\documentclass[twocolumn]{aastex62}
\usepackage{amssymb}
\usepackage{natbib}
\usepackage{hyperref}
\usepackage{color}
\usepackage{pdfcomment}

\submitjournal{AJ}
\received{2018 August 27}
\accepted{2018 September 26}

\shorttitle{Reactivations of 238P and 288P}
\shortauthors{Hsieh et al.}

\begin{document}

\title{The 2016 Reactivations of Main-Belt Comets 238P/Read and 288P/(300163) 2006 VW$_{139}$\footnote{Based on observations obtained with MegaPrime/MegaCam, a joint project of CFHT and CEA/DAPNIA, at the Canada-France-Hawaii Telescope (CFHT) (Programs 12BH43, 15AT05, and 16BT05), which is operated by the National Research Council (NRC) of Canada, the Institut National des Science de l'Univers of the Centre National de la Recherche Scientifique (CNRS) of France, and the University of Hawaii, and at the Gemini Observatory (Programs GN-2011B-Q-17, GN-2012A-Q-68, GN-2012B-Q-106, GN-2016B-LP-11, and GS-2016B-LP-11), which is operated by the Association of Universities for Research in Astronomy, Inc., under a cooperative agreement with the National Science Foundation (NSF) on behalf of the Gemini partnership: the NSF (United States), the National Research Council (Canada), CONICYT (Chile), Ministerio de Ciencia, Tecnolog\'ia e Innovaci\'on Productiva (Argentina), and Minist\'erio da Ci\^encia, Tecnologia e Inova\c{c}\=ao (Brazil).}}

\correspondingauthor{Henry H.\ Hsieh}
\email{hhsieh@psi.edu}

\author[0000-0001-7225-9271]{Henry H.\ Hsieh}
\affil{Planetary Science Institute, 1700 East Fort Lowell Rd., Suite 106, Tucson, AZ 85719, USA}
\affil{Institute of Astronomy and Astrophysics, Academia Sinica, P.O.\ Box 23-141, Taipei 10617, Taiwan}

\author[0000-0002-7332-2479]{Masateru Ishiguro}
\affil{Department of Physics and Astronomy, Seoul National University, Gwanak, Seoul 151-742, Korea}

\author[0000-0002-4676-2196]{Yoonyoung Kim}
\affil{Department of Physics and Astronomy, Seoul National University, Gwanak, Seoul 151-742, Korea}
\affil{Max Planck Institute for Solar System Research, Justus-von-Liebig-Weg 3, D-37077 G{\"o}ttingen, Germany}

\author[0000-0003-2781-6897]{Matthew M.\ Knight}
\affil{Department of Astronomy, University of Maryland, 1113 Physical Sciences Complex, Building 415, College Park, MD 20742, USA}

\author{Zhong-Yi Lin}
\affil{Institute of Astronomy, National Central University, Jhongli, Taiwan}

\author[0000-0001-7895-8209]{Marco Micheli}
\affil{ESA SSA-NEO Coordination Centre, Largo Galileo Galilei, 1, 00044 Frascati (RM), Italy}
\affil{INAF - Osservatorio Astronomico di Roma, Via Frascati, 33, 00040 Monte Porzio Catone (RM), Italy}

\author[0000-0001-6765-6336]{Nicholas A.\ Moskovitz}
\affil{Lowell Observatory, 1400 W.\ Mars Hill Rd, Flagstaff, AZ 86011, USA}

\author[0000-0003-3145-8682]{Scott S.\ Sheppard}
\affil{Department of Terrestrial Magnetism, Carnegie Institution for Science, 5241 Broad Branch Road NW, Washington, DC 20015, USA}

\author[0000-0002-1506-4248]{Audrey Thirouin}
\affil{Lowell Observatory, 1400 W.\ Mars Hill Rd, Flagstaff, AZ 86011, USA}

\author[0000-0001-9859-0894]{Chadwick A.\ Trujillo}
\affil{Department of Physics and Astronomy, Northern Arizona University, Flagstaff, AZ 86011, USA}



\begin{abstract} 
We report observations of the reactivations of main-belt comets 238P/Read and 288P/(300163) 2006 VW$_{139}$, that also track the evolution of each object's activity over several months in 2016 and 2017.  We additionally identify and analyze archival SDSS data showing 288P to be active in 2000, meaning that both 238P and 288P have now each been confirmed to be active near perihelion on three separate occasions.  From data obtained of 288P from 2012-2015 when it appeared inactive, we find best-fit $R$-band $H,G$ phase function parameters of $H_R=16.80\pm0.12$~mag and $G_R=0.18\pm0.11$, corresponding to effective component radii of $r_c=0.80\pm0.04$~km, assuming a binary system with equally-sized components.  Fitting linear functions to ejected dust masses inferred for 238P and 288P soon after their observed reactivations in 2016, we find an initial average net dust production rate of ${\dot M}_d=0.7\pm0.3$~kg~s$^{-1}$ and a best-fit start date of 2016 March 11 (when the object was at a true anomaly of $\nu=-63^{\circ}$) for 238P, and an initial average net dust production rate of ${\dot M}_d=5.6\pm0.7$~kg~s$^{-1}$ and a best-fit start date of 2016 August 5 (when the object was at $\nu=-27^{\circ}$) for 288P.  Applying similar analyses to archival data, we find similar start points for previous active episodes for both objects, suggesting that minimal mantle growth or ice recession occurred between the active episodes in question. Some changes in dust production rates between active episodes are detected, however.  More detailed dust modeling is suggested to further clarify the process of activity evolution in main-belt comets.
\end{abstract}


\bigskip
\section{INTRODUCTION}\label{section:introduction}
\subsection{Background}\label{section:background}

Main-belt comets \citep[MBCs;][]{hsieh2006_mbcs} are objects that orbit in the main asteroid belt but also exhibit comet-like activity that has been determined to at least be partially due to the sublimation of volatile ices, typically assumed to be water ice.  Exposed water ice is thermally unstable against sublimation at the temperatures found in the main asteroid belt, and thus is not expected to remain very long on the surface of main-belt asteroids.  Thermal modeling has shown, however, that buried water ice can survive within short distances of asteroid surfaces over the age of the solar system \citep{schorghofer2008_mbaice,schorghofer2016_asteroidice,prialnik2009_mbaice}, where it might then be susceptible to excavation and exposure to solar heating by small impacts \citep[e.g.,][]{hsieh2004_133p,hsieh2009_htp,capria2012_mbcactivity,haghighipour2016_mbcimpacts}.  While mantling and depletion of volatiles are expected to eventually extinguish activity triggered by an individual impact, an active MBC may nonetheless be expected to remain active over several orbit passages after its initial activation \citep[cf.][]{hsieh2015_ps1mbcs}.

MBCs are a subset of the active asteroids \citep[cf.][]{jewitt2015_actvasts_ast4}, which include all small bodies that exhibit comet-like activity yet occupy asteroid-like orbits. Small solar system bodies are commonly considered dynamically asteroidal if they have Tisserand parameter values (with respect to Jupiter) of $T_J>3$ \citep{kresak1979_cometasteroidinterrelations_ast1}, although in practice, the dynamical transition zone between asteroids and comets actually appears to lie roughly between $T_J=3.05$ and $T_J=3.10$ \citep{tancredi2014_asteroidcometclassification,jewitt2015_actvasts_ast4,hsieh2016_tisserand}.  Besides MBCs, the other major type of active asteroids are disrupted asteroids \citep[cf.][]{hsieh2012_scheila}, which exhibit comet-like activity due to effects other than sublimation, such as impact-driven dust ejection or rotational destabilization.

Although there is a non-zero chance that some MBCs (particularly those with both large eccentricities and large inclinations) may be implanted Jupiter-family comets (JFCs) originally from the outer solar system, most MBCs appear to be native to the main asteroid belt \citep{hsieh2016_tisserand}.  Thus, it may be possible to use MBCs to set temperature and compositional constraints on this region of the early solar system \citep{hsieh2016_mbcsiausproc}.  MBCs are intriguing because they offer new opportunities to constrain models of our solar system's formation and evolution \citep[cf.][]{hsieh2014_mbcsiausproc}, and also test hypotheses that icy objects from the region of the solar system currently occupied by the present-day main asteroid belt may have played a significant role in the primordial delivery of water and other volatiles to the early Earth \citep[][and references within]{morbidelli2000_earthwater,raymond2004_earthwater,raymond2017_waterorigin,obrien2006_earthwater,obrien2018_waterdelivery}.

\subsection{Recurrent Activity in MBCs}\label{section:recurrent}

Recurrent activity, especially occurring near perihelion with intervening periods of inactivity, is a strong indication that an active asteroid's activity is driven by sublimation, making it key to differentiating between MBCs and disrupted asteroids.  Repeated impacts or rotational destabilization events are not expected to occur on the same asteroid on timescales as short as a single main-belt asteroid orbit (i.e., 5-6 years), nor are they expected to exhibit periodic behavior correlated with an object's orbital position.  On the other hand, repeated activity near perihelion is naturally explained by sublimation-driven active behavior where such activity is correlated with the higher temperatures experienced by an object during perihelion passages \citep{hsieh2012_scheila}.

This conclusion is further strengthened by the trend noted by \citet{hsieh2015_324p} that the active episodes of all nine active asteroids identified as MBCs thus far via other analyses (e.g., dust modeling or photometric monitoring) all occur near perihelion (i.e., within true anomaly ranges of $-60^{\circ}\lesssim\nu\lesssim120^{\circ}$, where $\nu=0^{\circ}$ at perihelion).  This finding supports the idea that their active episodes are correlated with temperature, as would be expected if they were due to sublimation, and do not occur at arbitrary orbit positions, as would be expected if they were due to impact or rotational disruption.  To date, seven MBCs have been confirmed to be recurrently active: 133P/Elst-Pizarro \citep{elst1996_133p,hsieh2004_133p}, 238P/Read \citep{hsieh2009_238p,hsieh2011_238p}, 288P \citep{hsieh2012_288p,agarwal2016_288p_cbet}, 259P/Garradd \citep{jewitt2009_259p,hsieh2017_259p}, 313P/Gibbs \citep{hsieh2015_313p,hui2015_313p}, 324P/La Sagra \citep{hsieh2012_324p,hsieh2015_324p}, and 358P/PANSTARRS \citep{hsieh2018_358p}.

Searching for recurrent activity is also important because most MBCs are discovered while they are already active, and even those that were already known as inactive asteroids before being discovered to be active (e.g., 133P, 176P, 288P) were not being regularly monitored leading up to the discovery of their activity.  As such, the onset times of MBC active periods are generally poorly constrained.  By performing an observational monitoring campaign starting well in advance of the expected return of activity in a MBC candidate, more explicit constraints can be placed on the onset time of that activity. Coupled with analogous observations of the decline of activity, constraints on the onset time of activity can in turn allow stronger constraints to be placed on physical quantities of interest such as the total duration of activity, total mass loss per active episode, and the depth of the buried ice presumed to drive the observed activity.  Characterizing the evolution of activity strength over multiple active episodes may also help constrain total active lifetimes following each triggering event on a MBC, which in turn may give greater context to discovery statistics \citep[cf.\ Section~\ref{section:discussion};][]{hsieh2009_htp,hsieh2015_ps1mbcs}.

Given the above considerations, we have conducted observational campaigns to search for and characterize the return of activity for MBCs 238P and 288P, and report the results in this work.


\subsection{238P/Read}\label{section:intro_238p}

Comet 238P/Read (formerly designated P/2005 U1) was discovered to be an active comet on UT 2005 October 24 \citep{read2005_238p}, shortly after perihelion.  
It was one of the first three objects that led to the recognition of MBCs as a new class of comets \citep{hsieh2006_mbcs}.  The absolute magnitude of the object's nucleus has been measured to be $H_R$$\,=\,$19.05$\pm$0.05~mag, corresponding to an effective nucleus radius of $r_n$$\,\sim\,$0.4~km, assuming an albedo of $p_R$$\,=\,$0.05 \citep{hsieh2011_238p}.

An initial analysis of 238P's activity employing Finson-Probstein-style dust modeling \citep{finson1968_cometdustmodeling1} showed that it was most likely driven by ice sublimation \citep{hsieh2009_238p}.  This conclusion was strongly supported by the 2010 appearance of recurrent activity leading up to the object's next perihelion passage in 2011 \citep{hsieh2011_238p}.  Dust modeling of the 2005 active episode indicated that the coma and tail were optically dominated by dust particles $>$10~$\mu$m in radius with terminal ejection velocities of 0.2-3~m~s$^{-1}$, and that the object's average mass loss rate during the period of observation was approximately 0.2~kg~s$^{-1}$ \citep{hsieh2009_238p}.  A photometric analysis showed indications that the object's activity in 2010-2011 may have been comparable in strength to the activity observed in 2005 \citep{hsieh2011_238p}. However, the observing periods being compared in that analysis did not actually overlap, and as such, no definitive conclusions could be drawn at the time from the available data about the change (or lack thereof) in activity strength from the first epoch to the next.

Notably, 238P was also the primary target of the proposed NASA Discovery mission, Proteus, to visit and physically characterize a MBC and its activity \citep{meech2015_proteus}.

\subsection{288P/(300163) 2006 VW$_{139}$}\label{section:intro_288p}

Comet 288P is also known by its asteroid designation, (300163) 2006 VW$_{139}$, by which it was already known at the time of the discovery of its activity in 2011 \citep{hsieh2012_288p}.  It was the first of now numerous active asteroids to be discovered to date by the Pan-STARRS1 (PS1) survey telescope \citep{chambers2016_panstarrs}.  Following its discovery, deep follow-up observations showed a short antisolar dust tail and a longer dust trail aligned with the object's orbit plane, indicative of the simultaneous presence of recent dust emission (in the antisolar dust tail) and older dust emission (in the orbit-plane-aligned dust trail), and therefore of a prolonged dust emission event.  A photometric analysis showed the object maintaining a roughly constant coma brightness for about a month, also suggesting the action of a sustained dust emission event, characteristic of sublimation-driven activity, and not, for instance, of an impulsive driver such as an impact.  Spectroscopic searches for CN emission were unsuccessful, although did provide upper limit production rates.
\citet{hsieh2012_288p} found $Q_{\rm CN}$$\,<\,$1.3$\times$10$^{24}$~molecules~s$^{-1}$, roughly equivalent to an upper limit water production rate of $Q_{\rm H_2O}$$\,<\,$10$^{26}$~molecules~s$^{-1}$, assuming Jupiter-family comet-like chemical composition ratios, which we note is not necessarily a valid assumption for MBCs.  Meanwhile, \citep{licandro2013_288p} found $Q_{\rm CN}$$\,<\,$3.8$\times$10$^{23}$~molecules~s$^{-1}$.

Dust modeling analyses of the object's 2011 activity found an average dust production rate of $\sim\,$0.2~kg~s$^{-1}$ and typical ejection velocities of $\sim\,$0.1-0.3~m~s$^{-1}$ \citep{licandro2013_288p,agarwal2016_288p}.  At the time, the best estimate of the absolute magnitude of the object's nucleus was $H_R$$\,=\,$16.4~mag, corresponding to an effective nucleus radius of $r_n$$\,\sim\,$1.4~km, assuming an albedo of $p_R$$\,=\,$0.05 \citep{hsieh2012_288p}. A later analysis of {\it Hubble Space Telescope} data indicated that the nucleus had an absolute magnitude of $H_V$$\,=\,$17.0$\pm$0.1~mag \citep{agarwal2016_288p}, equivalent to $H_R$$\,\sim\,$16.6~mag, assuming solar colors.  The nucleus has been spectroscopically classified as a C-type asteroid \citep{licandro2013_288p}, and has been identified as a binary system \citep{agarwal2017_288p}.

Interestingly, 288P was found to potentially belong to a small, young 11-member cluster of asteroids just 7.5$\pm$0.3~Myr old \citep{novakovic2012_288p}.  Together with the discovery that fellow MBC 133P belonged to the $<$10~Myr-old Beagle family \citep{nesvorny2008_beagle}, this finding suggests that the younger (and therefore potentially more volatile-rich) surfaces of members of these families could be more susceptible to activation by small impactors excavating shallow buried ice and thus be more likely to develop observable activity \citep[e.g.,][]{hsieh2018_activeastfamilies}.  As such, \citet{novakovic2012_288p} hypothesized both that currently known young families could be productive regions in which to search for new MBCs, and that currently known MBCs might be found in the future to be part of as-yet undiscovered young families.

\setlength{\tabcolsep}{2.5pt}
\setlength{\extrarowheight}{0em}
\begin{table*}[htb]
\caption{Observation Log: 238P - Active}
\centering
\smallskip
\footnotesize
\begin{tabular}{lccrcrccrcccrr}
\hline\hline
\multicolumn{1}{c}{UT Date}
 & \multicolumn{1}{c}{Tel.$^a$}
 & \multicolumn{1}{c}{$N$$^b$}
 & \multicolumn{1}{c}{$t$$^c$}
 & \multicolumn{1}{c}{Filter}
 & \multicolumn{1}{c}{$\nu$$^d$}
 & \multicolumn{1}{c}{$R$$^e$}
 & \multicolumn{1}{c}{$\Delta$$^f$}
 & \multicolumn{1}{c}{$\alpha$$^g$}
 & \multicolumn{1}{c}{$m_{R,n}$$^h$}
 & \multicolumn{1}{c}{$m_{R,t}$$^i$}
 & \multicolumn{1}{c}{$H_{R,t}$$^j$}
 & \multicolumn{1}{c}{$M_d$$^k$}
 & \multicolumn{1}{c}{$Af\rho$$^l$}
 \\
\hline
2011 Mar 10 & \multicolumn{4}{l}{\it Perihelion .................} &   0.0 & 2.361 & 3.276 &  8.1 & --- & --- & --- & \multicolumn{1}{c}{---} & \multicolumn{1}{c}{---} \\
2011 Aug 29 & Gemini-N &  2 &  360 & $r'$ &  49.8 & 2.543 & 2.751 & 21.5 & 21.97$\pm$0.05 & 21.9$\pm$0.1 & 16.4$\pm$0.1 & (1.8$\pm$0.2)$\times$10$^7$ &  4.2$\pm$0.6 \\
2011 Sep 25 & Gemini-N &  5 &  900 & $r'$ &  56.7 & 2.598 & 2.471 & 22.7 & 22.25$\pm$0.05 & 21.9$\pm$0.1 & 16.5$\pm$0.1 & (1.6$\pm$0.2)$\times$10$^7$ &  2.9$\pm$0.4 \\
2011 Dec 31 & Gemini-N &  9 & 1620 & $r'$ &  79.2 & 2.825 & 1.844 &  1.8 & 21.54$\pm$0.05 & 21.2$\pm$0.1 & 17.3$\pm$0.1 & (0.6$\pm$0.1)$\times$10$^7$ &  1.1$\pm$0.2 \\
2016 Jul  8 & CFHT     &  5 &  900 & $r'$ & $-$31.5 & 2.439 & 2.095 & 24.4 & 22.32$\pm$0.05 & 22.3$\pm$0.1 & 17.3$\pm$0.1 & (0.7$\pm$0.1)$\times$10$^7$ &  2.0$\pm$0.3 \\ 
2016 Aug  6 & Gemini-N &  5 &  900 & $r'$ & $-$23.1 & 2.405 & 1.742 & 21.6 & 21.60$\pm$0.05 & 21.5$\pm$0.1 & 17.1$\pm$0.1 & (0.9$\pm$0.1)$\times$10$^7$ &  3.0$\pm$0.4 \\ 
2016 Sep  5 & Gemini-N &  5 &  900 & $r'$ & $-$14.2 & 2.381 & 1.467 & 13.1 & 20.69$\pm$0.05 & 20.3$\pm$0.1 & 16.6$\pm$0.1 & (1.4$\pm$0.2)$\times$10$^7$ &  4.1$\pm$0.5 \\ 
2016 Sep  6 & CFHT     &  5 &  900 & $r'$ & $-$13.9 & 2.380 & 1.461 & 12.7 & 20.76$\pm$0.05 & 20.3$\pm$0.1 & 16.7$\pm$0.1 & (1.4$\pm$0.2)$\times$10$^7$ &  3.7$\pm$0.5 \\ 
2016 Sep 23 & LOT      &  5 & 1500 & $r'$ &  $-$8.8 & 2.372 & 1.382 &  5.1 & 20.14$\pm$0.05 & 20.0$\pm$0.1 & 16.9$\pm$0.1 & (1.1$\pm$0.1)$\times$10$^7$ &  4.3$\pm$0.5 \\ 
2016 Sep 25 & CFHT     &  5 &  900 & $r'$ &  $-$8.2 & 2.371 & 1.377 &  4.3 & 20.06$\pm$0.05 & 19.7$\pm$0.1 & 16.7$\pm$0.1 & (1.4$\pm$0.1)$\times$10$^7$ &  4.4$\pm$0.5 \\ 
2016 Sep 26 & CFHT     &  5 &  900 & $r'$ &  $-$7.9 & 2.371 & 1.375 &  3.8 & 20.05$\pm$0.05 & 19.7$\pm$0.1 & 16.7$\pm$0.1 & (1.3$\pm$0.1)$\times$10$^7$ &  4.3$\pm$0.5 \\ 
2016 Sep 27 & Gemini-N &  5 &  875 & $r'$ &  $-$7.6 & 2.370 & 1.374 &  3.3 & 20.03$\pm$0.05 & 19.7$\pm$0.1 & 16.7$\pm$0.1 & (1.2$\pm$0.1)$\times$10$^7$ &  4.2$\pm$0.5 \\ 
2016 Oct  3 & LOT      & 14 & 4200 & $r'$ &  $-$5.7 & 2.369 & 1.368 &  0.6 & 19.79$\pm$0.05 & 19.7$\pm$0.1 & 17.0$\pm$0.1 & (0.9$\pm$0.1)$\times$10$^7$ &  4.0$\pm$0.4 \\ 
2016 Oct  9 & DCT      &  7 & 1500 & $R$  &  $-$4.0 & 2.367 & 1.372 &  2.8 & 19.83$\pm$0.05 & 19.4$\pm$0.1 & 16.5$\pm$0.1 & (1.6$\pm$0.2)$\times$10$^7$ &  5.0$\pm$0.5 \\ 
2016 Oct 22 & \multicolumn{4}{l}{\it Perihelion .................} &   0.0 & 2.366 & 1.415 &  9.1 & --- & --- & --- & \multicolumn{1}{c}{---} & \multicolumn{1}{c}{---} \\
2016 Oct 25 & LOT      &  5 & 1500 & $r'$ &   0.9 & 2.366 & 1.428 & 10.3 & 19.83$\pm$0.05 & 19.7$\pm$0.1 & 16.2$\pm$0.1 & (2.1$\pm$0.2)$\times$10$^7$ &  8.4$\pm$0.9 \\
2016 Nov  5 & CFHT     &  3 &  540 & $r'$ &   4.2 & 2.367 & 1.499 & 14.6 & 19.97$\pm$0.05 & 19.9$\pm$0.1 & 16.1$\pm$0.1 & (2.3$\pm$0.3)$\times$10$^7$ &  9.5$\pm$1.1 \\
2016 Dec 22 & Gemini-N &  3 &  540 & $r'$ &  18.3 & 2.390 & 2.009 & 23.9 & 20.70$\pm$0.05 & 20.5$\pm$0.1 & 15.7$\pm$0.1 & (3.6$\pm$0.5)$\times$10$^7$ &  9.6$\pm$1.3 \\
2016 Dec 26 & CFHT     &  7 & 1260 & $r'$ &  19.5 & 2.394 & 2.061 & 24.0 & 20.73$\pm$0.05 & 20.5$\pm$0.1 & 15.6$\pm$0.1 & (3.8$\pm$0.5)$\times$10$^7$ &  9.7$\pm$1.3 \\
2016 Dec 29 & Gemini-N &  5 &  900 & $r'$ &  20.3 & 2.396 & 2.100 & 24.1 & 20.64$\pm$0.05 & 20.4$\pm$0.1 & 15.5$\pm$0.1 & (4.4$\pm$0.6)$\times$10$^7$ & 10.9$\pm$1.4 \\
2016 Dec 29 & CFHT     &  4 &  720 & $r'$ &  20.3 & 2.396 & 2.100 & 24.1 & 20.48$\pm$0.05 & 20.4$\pm$0.1 & 15.5$\pm$0.1 & (4.4$\pm$0.6)$\times$10$^7$ & 12.7$\pm$1.7 \\
2017 Jan 18 & Gemini-N &  1 &  180 & $r'$ &  26.2 & 2.416 & 2.367 & 23.7 & 20.70$\pm$0.05 & 20.6$\pm$0.1 & 15.4$\pm$0.1 & (4.6$\pm$0.6)$\times$10$^7$ & 11.7$\pm$1.5 \\
2017 Jan 26 & Gemini-N &  5 &  900 & $r'$ &  28.5 & 2.425 & 2.473 & 23.2 & 20.99$\pm$0.05 & 20.7$\pm$0.1 & 15.4$\pm$0.1 & (4.6$\pm$0.6)$\times$10$^7$ &  9.2$\pm$1.2 \\
2022 Jun  5 & \multicolumn{4}{l}{\it Perihelion .................} &   0.0 & 2.369 & 3.024 & 16.7 & --- & --- & --- & \multicolumn{1}{c}{---} & \multicolumn{1}{c}{---} \\
\hline
\hline
\multicolumn{14}{l}{$^a$ Telescope used.} \\
\multicolumn{14}{l}{$^b$ Number of exposures.} \\
\multicolumn{14}{l}{$^c$ Total integration time, in seconds.} \\
\multicolumn{14}{l}{$^d$ True anomaly, in degrees.} \\
\multicolumn{14}{l}{$^e$ Heliocentric distance, in au.} \\
\multicolumn{14}{l}{$^f$ Geocentric distance, in au.} \\
\multicolumn{14}{l}{$^g$ Solar phase angle (Sun-object-Earth), in degrees.} \\
\multicolumn{14}{l}{$^h$ Equivalent mean apparent $R$-band nucleus magnitude, measured within photometry apertures with radii of $4\farcs0$.} \\
\multicolumn{14}{l}{$^i$ Equivalent total mean apparent $R$-band magnitude, including the entire coma and tail, if present.} \\
\multicolumn{14}{l}{$^j$ Total absolute $R$-band magnitude, using $H,G$ phase function where $G=-0.03$.} \\
\multicolumn{14}{l}{$^k$ Estimated total dust mass, in kg, assuming $\rho_d\sim2500$~kg~m$^3$.} \\
\multicolumn{14}{l}{$^l$ $Af\rho$ values computed using photometry apertures with radii of $4\farcs0$, in cm, where uncertainties are estimated to be $\sim$10\%.} \\
\end{tabular}
\label{table:obs_238p_active}
\end{table*}

\setlength{\tabcolsep}{5pt}
\begin{table*}[htb!]
\caption{Previously reported observations of 238P when active$^a$}
\centering
\smallskip
\footnotesize
\begin{tabular}{lcrccrcccr}
\hline\hline
\multicolumn{1}{c}{UT Date}
 & \multicolumn{1}{c}{Tel.$^b$}
 & \multicolumn{1}{c}{$\nu$$^c$}
 & \multicolumn{1}{c}{$R$$^d$}
 & \multicolumn{1}{c}{$\Delta$$^e$}
 & \multicolumn{1}{c}{$\alpha$$^f$}
 & \multicolumn{1}{c}{$m_{R,n}$$^g$}
 & \multicolumn{1}{c}{$m_{R,t}$$^h$}
 & \multicolumn{1}{c}{$H_{R,t}$$^i$}
 & \multicolumn{1}{c}{$M_d$$^j$}
 \\
\hline
2005 Jul 27 & \multicolumn{1}{l}{\it Perihelion ...} &   0.0 & 2.365 & 2.276 & 25.2 & --- & --- & --- & \multicolumn{1}{c}{---} \\
2005 Nov 10 & UH2.2    &    31.4 & 2.437 & 1.447 &  0.5 & 19.28$\pm$0.05 & $<$19.3 & $<$16.4 & $>$(1.7$\pm$0.2)$\times$10$^7$ \\ 
2005 Nov 19 & UH2.2    &    33.9 & 2.449 & 1.469 &  3.8 & 19.34$\pm$0.05 & $<$19.3 & $<$16.1 & $>$(2.4$\pm$0.2)$\times$10$^7$ \\ 
2005 Nov 20 & UH2.2    &    34.2 & 2.450 & 1.473 &  4.3 & 19.46$\pm$0.05 & $<$19.5 & $<$16.0 & $>$(2.4$\pm$0.3)$\times$10$^7$ \\ 
2005 Nov 21 & UH2.2    &    34.5 & 2.452 & 1.477 &  7.3 & 19.37$\pm$0.05 & $<$19.4 & $<$16.1 & $>$(2.4$\pm$0.2)$\times$10$^7$ \\ 
2005 Nov 22 & UH2.2    &    34.8 & 2.453 & 1.481 &  7.8 & 19.28$\pm$0.05 & $<$19.3 & $<$15.9 & $>$(2.7$\pm$0.3)$\times$10$^7$ \\ 
2005 Nov 26 & Gemini-N &    35.9 & 2.459 & 1.501 &  9.5 & 19.72$\pm$0.05 & $<$19.7 & $<$16.2 & $>$(2.1$\pm$0.2)$\times$10$^7$ \\ 
2005 Dec 24 & UH2.2    &    43.6 & 2.504 & 1.743 & 17.1 & 20.12$\pm$0.05 & $<$20.1 & $<$15.8 & $>$(3.3$\pm$0.4)$\times$10$^7$ \\ 
2005 Dec 25 & UH2.2    &    43.9 & 2.506 & 1.754 & 19.9 & 20.16$\pm$0.05 & $<$20.2 & $<$15.8 & $>$(3.1$\pm$0.4)$\times$10$^7$ \\ 
2010 Sep  3 & UH2.2    & $-$54.1 & 2.576 & 1.643 & 10.7 & 22.0$\pm$0.1   & $<$22.0 & $<$18.0 & $>$(0.3$\pm$0.1)$\times$10$^7$ \\ 
2010 Sep  4 & NTT      & $-$53.9 & 2.574 & 1.647 & 11.0 & 22.3$\pm$0.1   & $<$22.3 & $<$18.3 & $>$(0.2$\pm$0.1)$\times$10$^7$ \\ 
2010 Sep  5 & NTT      & $-$53.6 & 2.572 & 1.651 & 11.4 & 22.3$\pm$0.1   & $<$22.3 & $<$18.3 & $>$(0.2$\pm$0.1)$\times$10$^7$ \\ 
2010 Oct  5 & Keck     & $-$45.7 & 2.514 & 1.869 & 20.3 & 22.3$\pm$0.1   & $<$22.3 & $<$17.7 & $>$(0.4$\pm$0.1)$\times$10$^7$ \\ 
2010 Nov 25 & UH2.2    & $-$31.5 & 2.433 & 2.414 & 23.5 & 21.8$\pm$0.1   & $<$21.8 & $<$16.6 & $>$(1.5$\pm$0.2)$\times$10$^7$ \\ 
2010 Dec  9 & UH2.2    & $-$27.5 & 2.416 & 2.566 & 22.5 & 21.9$\pm$0.1   & $<$21.9 & $<$16.6 & $>$(1.5$\pm$0.2)$\times$10$^7$ \\ 
2011 Mar 10 & \multicolumn{1}{l}{\it Perihelion ...} &   0.0 & 2.361 & 3.276 &  8.1 & --- & --- & --- & \multicolumn{1}{c}{---} \\
\hline
\hline
\multicolumn{10}{l}{$^a$ All 2005 data from \citet{hsieh2009_238p}, and all 2010 data from \citet{hsieh2011_238p}.} \\
\multicolumn{10}{l}{$^b$ Telescope (UH2.2: UH 2.2~m telescope; NTT: New Technology Telescope; Keck: Keck I Observatory).} \\
\multicolumn{10}{l}{$^c$ True anomaly, in degrees.} \\
\multicolumn{10}{l}{$^d$ Heliocentric distance, in au.} \\
\multicolumn{10}{l}{$^e$ Geocentric distance, in au.} \\
\multicolumn{10}{l}{$^f$ Solar phase angle (Sun-object-Earth), in degrees.} \\
\multicolumn{10}{l}{$^g$ Reported mean apparent $R$-band nucleus magnitude.} \\
\multicolumn{10}{l}{$^h$ Equivalent total apparent $R$-band magnitude, including the entire coma and tail, if present.} \\
\multicolumn{10}{l}{$^i$ Total absolute $R$-band magnitude, using $H,G$ phase function, where $G=-0.03$.} \\
\multicolumn{10}{l}{$^j$ Estimated total dust mass, in kg, assuming $\rho_d\sim2500$~kg~m$^3$.} \\
\end{tabular}
\label{table:obs_238p_active_prev}
\end{table*}

Observations obtained with the {\it Hubble Space Telescope} on 2016 August 22 revealed that 288P had again become active following its observed 2011-2012 active episode \citep{agarwal2016_288p_cbet}.  With these observations, 288P became the fifth MBC to be confirmed to exhibit recurrent activity, after 133P, 238P, 313P, and 324P. Since these observations, recurrent activity in 259P and 358P has also been confirmed, bringing the current total of MBCs confirmed to exhibit recurrent activity to seven.

\section{Observations}\label{section:observations}

\begin{figure}[htb!]
\centerline{\includegraphics[width=3.0in]{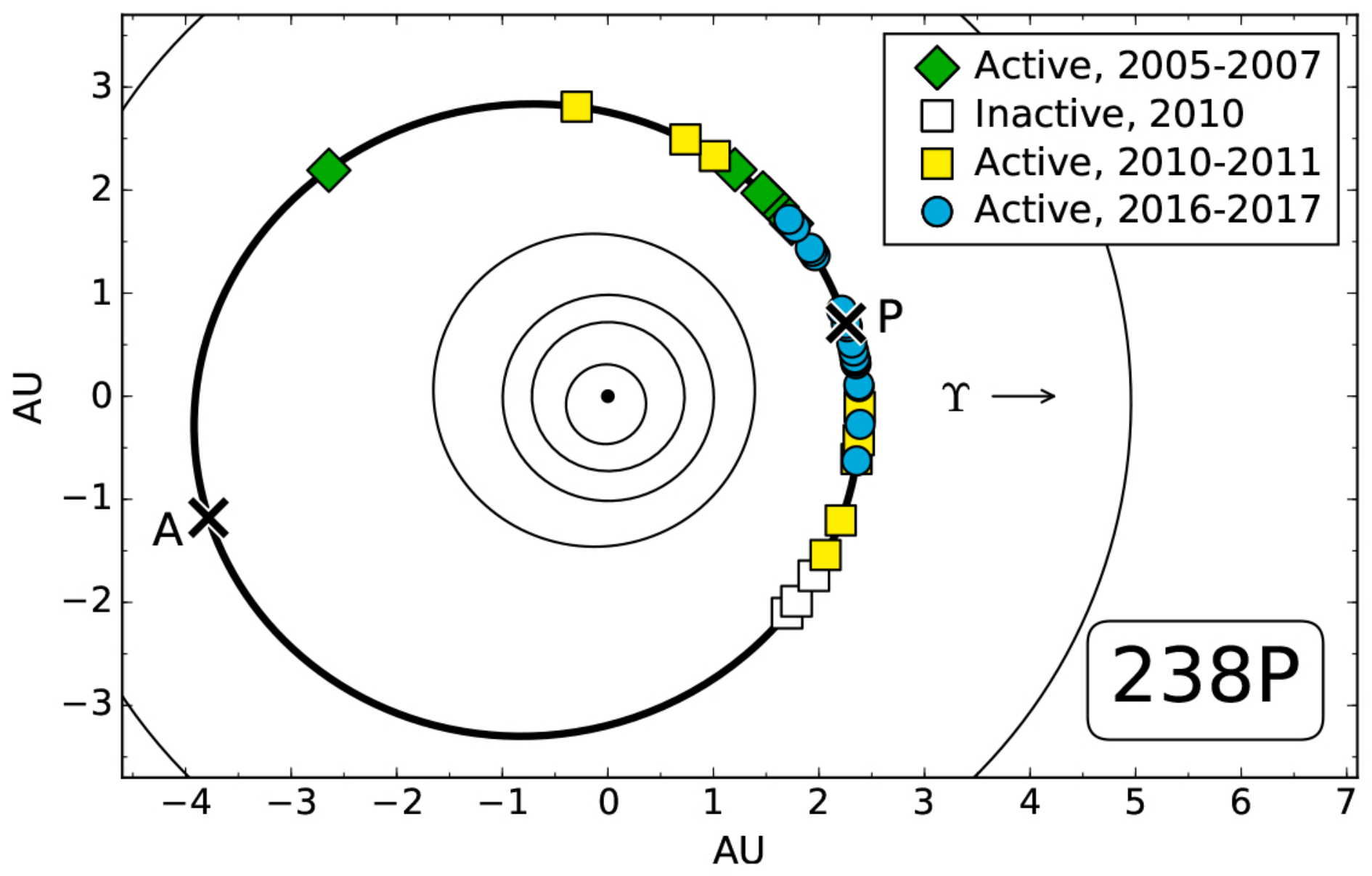}}
\caption{\small Orbit position plot with the Sun (black dot) at the center, the orbits of Mercury, Venus, Earth, Mars, 238P, and Jupiter marked by thin black lines, and the orbit of 238P marked by a thick black line. Perihelion (P) and aphelion (A) are marked with crosses. Green diamonds mark positions of observations when 238P was active in 2005-2007, open squares mark positions of observations when 238P was apparently inactive in 2010, yellow squares mark positions of observations when 238P was active in 2010-2011, and blue circles mark positions of observations when 238P was active in 2016-2017.
}
\label{figure:orbit_238p}
\end{figure}

\begin{figure*}[htb!]
\centerline{\includegraphics[width=6.5in]{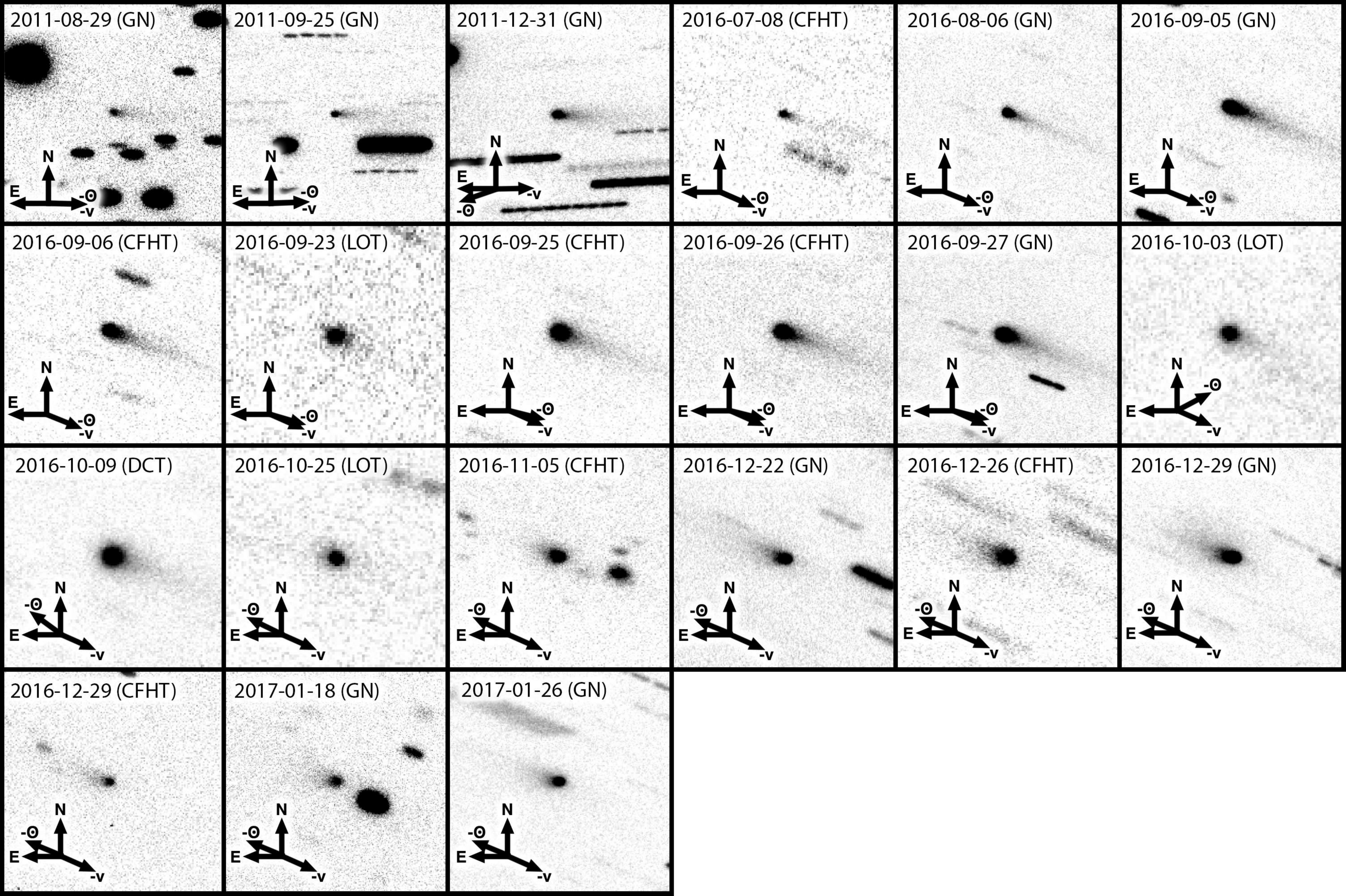}}
\caption{\small Composite $R$-band or $r'$-band images of 238P (at the center of each panel) constructed from data listed in Table~\ref{table:obs_238p_active}.  All panels are $30''\times30''$ in size, with north (N), east (E), the antisolar direction ($-\odot$), and the negative heliocentric velocity vector ($-v$), as projected on the sky, marked.  Panels are also labeled with dates of observations in YYYY-MM-DD format, as well as the telescope used to obtain each observation (CFHT: Canada-France-Hawaii Telescope; DCT: Discovery Channel Telescope; GN: Gemini-N; LOT: Lulin One-meter Telescope).
}
\label{figure:images_238p_active}
\end{figure*}

\begin{figure}[htb!]
\centerline{\includegraphics[width=3.0in]{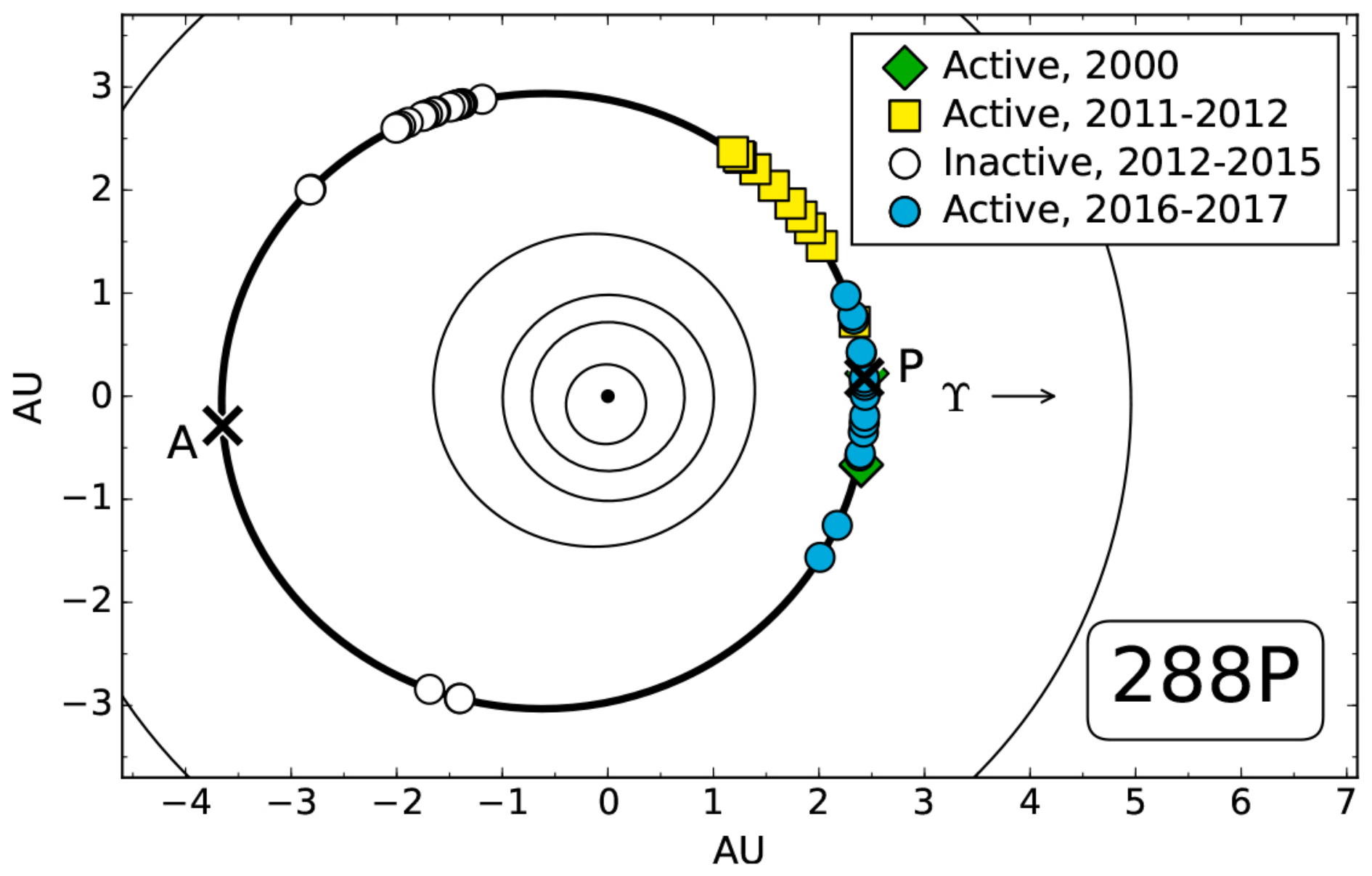}}
\caption{\small Orbit position plot with the Sun (black dot) at the center, the orbits of Mercury, Venus, Earth, Mars, 238P, and Jupiter marked by thin black lines, and the orbit of 288P marked by a thick black line. Perihelion (P) and aphelion (A) are marked with crosses. Green diamonds mark positions of observations when 288P was active in 2000, yellow squares mark positions of observations when 288P was active in 2011-2012, open circles mark positions of observations when 288P was apparently inactive in 2012-2015, and blue circles mark positions of observations when 288P was active in 2016-2017.
}
\label{figure:orbit_288p}
\end{figure}

\begin{figure*}[htb!]
\centerline{\includegraphics[width=6.5in]{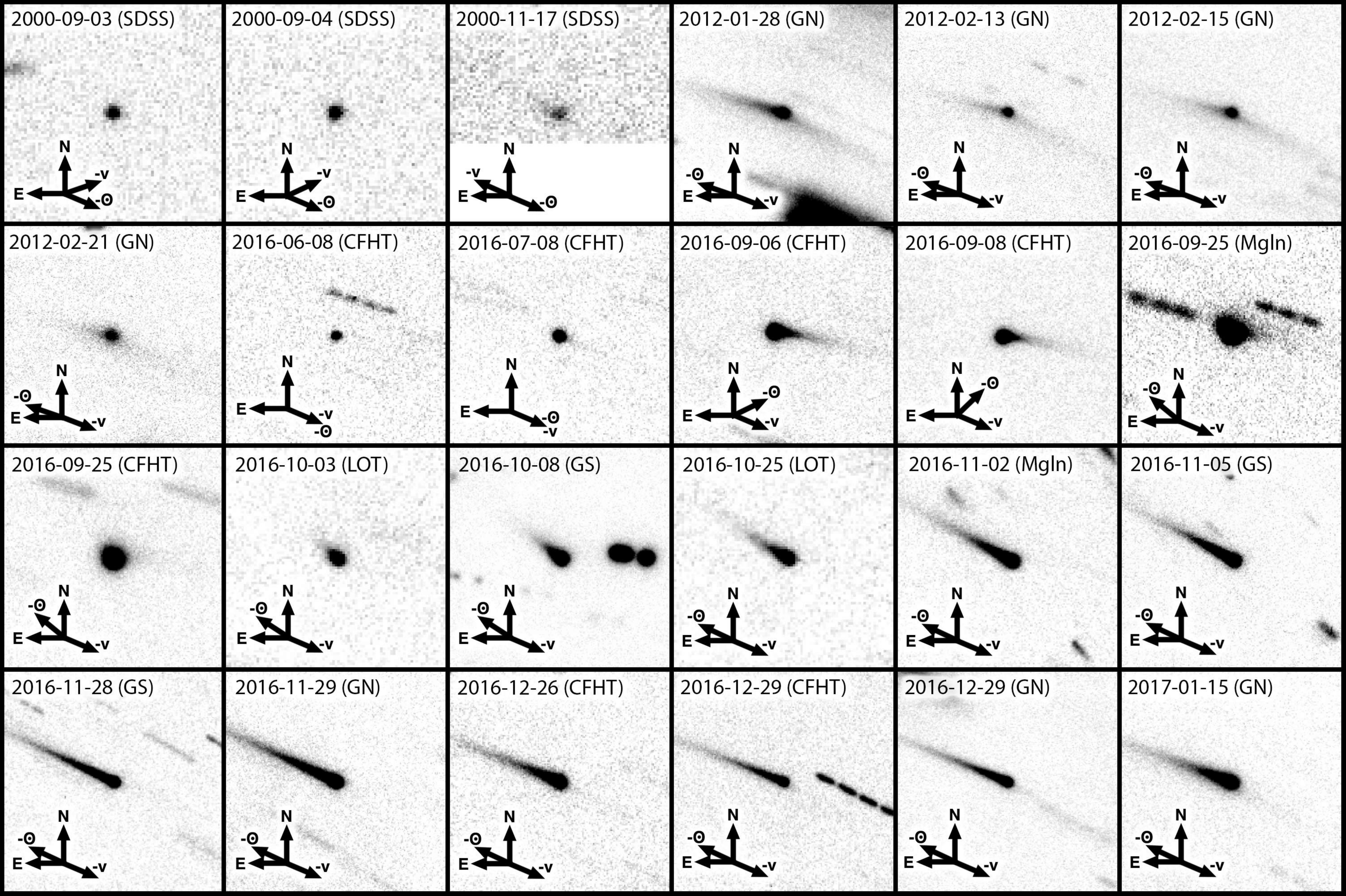}}
\caption{\small Composite $R$-band or $r'$-band images of 288P (at the center of each panel) constructed from data listed in Table~\ref{table:obs_288p_active}.  All panels are $30''\times30''$ in size, with north (N), east (E), the antisolar direction ($-\odot$), and the negative heliocentric velocity vector ($-v$), as projected on the sky, marked.
Panels are also labeled with dates of observations in YYYY-MM-DD format, as well as the telescope used to obtain each observation (CFHT: Canada-France-Hawaii Telescope; GN: Gemini-N; GS: Gemini-S; LOT: Lulin One-meter Telescope; Mgln: Magellan Baade telescope; SDSS: Sloan Digital Sky Survey telescope).
}
\label{figure:images_288p_active}
\end{figure*}

\begin{figure*}[htb!]
\centerline{\includegraphics[width=6.5in]{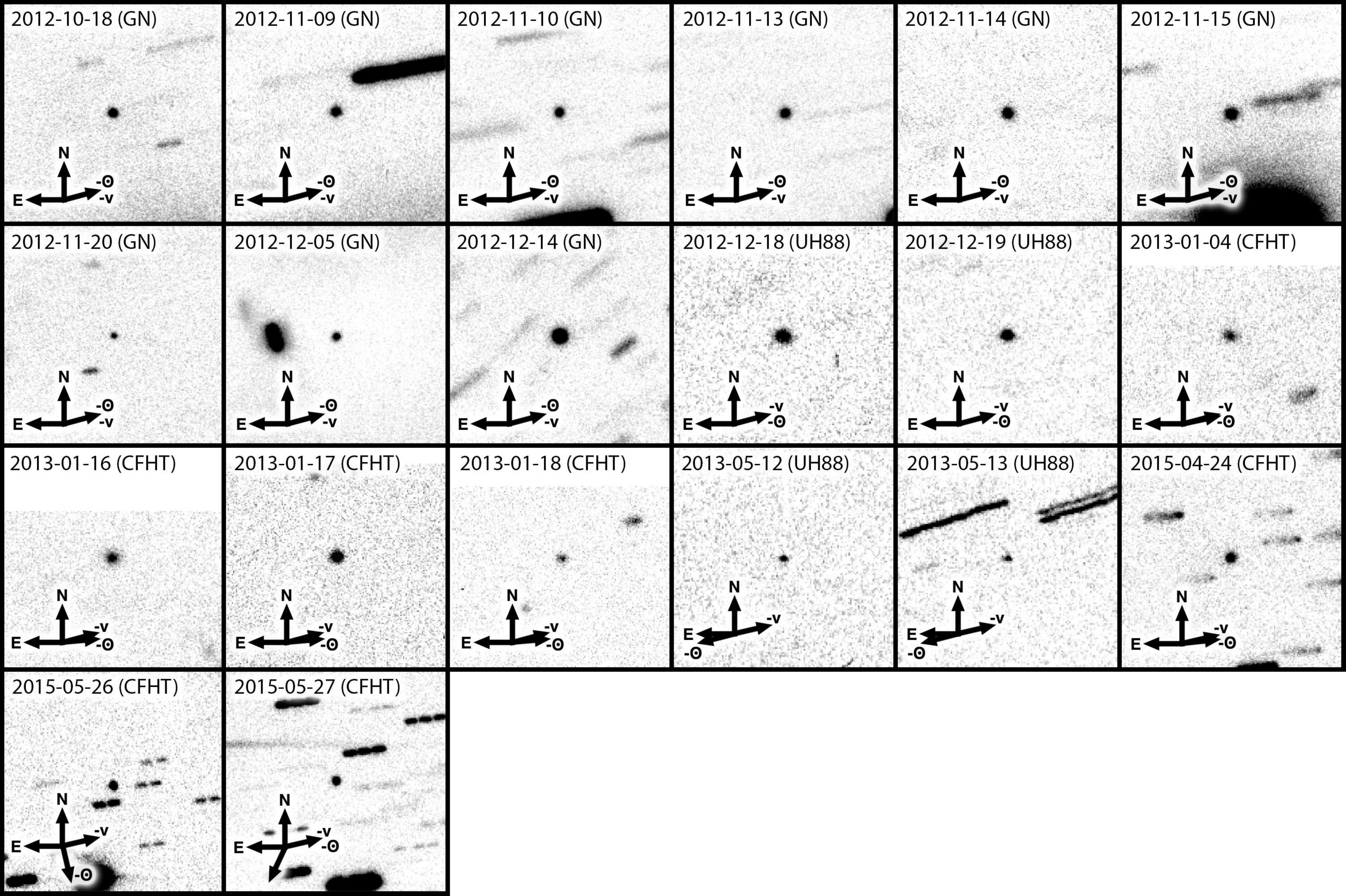}}
\caption{\small Composite $R$-band or $r'$-band images of 288P (at the center of each panel) constructed from data listed in Table~\ref{table:obs_288p_inactive}.  All panels are $30''\times30''$ in size, with north (N), east (E), the antisolar direction ($-\odot$), and the negative heliocentric velocity vector ($-v$), as projected on the sky, marked.
Panels are also labeled with dates of observations in YYYY-MM-DD format, as well as the telescope used to obtain each observation (CFHT: Canada-France-Hawaii Telescope; GN: Gemini-N; UH88: UH 2.2~m telescope).
}
\label{figure:images_288p_inactive}
\end{figure*}

\setlength{\tabcolsep}{2.5pt}
\setlength{\extrarowheight}{0em}
\begin{table*}[htb!]
\caption{Observation Log: 288P - Active}
\centering
\smallskip
\footnotesize
\begin{tabular}{lcrrcrccrcccrr}
\hline\hline
\multicolumn{1}{c}{UT Date}
 & \multicolumn{1}{c}{Tel.$^a$}
 & \multicolumn{1}{c}{$N$$^b$}
 & \multicolumn{1}{c}{$t$$^c$}
 & \multicolumn{1}{c}{Filter}
 & \multicolumn{1}{c}{$\nu$$^d$}
 & \multicolumn{1}{c}{$R$$^e$}
 & \multicolumn{1}{c}{$\Delta$$^f$}
 & \multicolumn{1}{c}{$\alpha$$^g$}
 & \multicolumn{1}{c}{$m_{R,n}$$^h$}
 & \multicolumn{1}{c}{$m_{R,t}$$^i$}
 & \multicolumn{1}{c}{$H_{R,t}$$^j$}
 & \multicolumn{1}{c}{$M_d$$^k$}
 & \multicolumn{1}{c}{$Af\rho$$^l$}
 \\
\hline
1995 Jul 13 & \multicolumn{4}{l}{\it Perihelion .......................} & 0.0 & 2.469 & 2.409 & 24.0 & --- & --- & --- & \multicolumn{1}{c}{---} & \multicolumn{1}{c}{---} \\
2000 Sep  3 & SDSS     &  1 &    54 & $r'$  & $-$21.1 & 2.491 & 1.488 &  3.1 & 19.48$\pm$0.05 & 19.5$\pm$0.1 & 16.4$\pm$0.1 & (0.7$\pm$0.3)$\times$10$^7$ &  3.2$\pm$1.1 \\ 
2000 Sep  4 & SDSS     &  1 &    54 & $r'$  & $-$20.8 & 2.490 & 1.486 &  2.8 & 19.36$\pm$0.05 & 19.4$\pm$0.1 & 16.3$\pm$0.1 & (0.8$\pm$0.3)$\times$10$^7$ &  4.0$\pm$1.2 \\ 
2000 Nov 17 & SDSS     &  1 &    54 & $r'$  &  $-$0.4 & 2.464 & 1.980 & 22.5 & 20.05$\pm$0.05 & 20.0$\pm$0.1 & 15.5$\pm$0.2 & (3.0$\pm$0.7)$\times$10$^7$ &  9.3$\pm$2.3 \\ 
2000 Nov 18 & \multicolumn{4}{l}{\it Perihelion .......................} & 0.0 & 2.464 & 1.997 & 22.6 & --- & --- & --- & \multicolumn{1}{c}{---} & \multicolumn{1}{c}{---} \\
2006 Mar 21 & \multicolumn{4}{l}{\it Perihelion .......................} & 0.0 & 2.445 & 3.437 &  1.7 & --- & --- & --- & \multicolumn{1}{c}{---} & \multicolumn{1}{c}{---} \\
2011 Jul 18 & \multicolumn{4}{l}{\it Perihelion .......................} & 0.0 & 2.438 & 2.292 & 24.6 & --- & --- & --- & \multicolumn{1}{c}{---} & \multicolumn{1}{c}{---} \\
2012 Jan 28 & Gemini-N &  7 &  1260 & $r'$  &    52.6 & 2.609 & 2.453 & 22.2 & 21.02$\pm$0.05 & 20.3$\pm$0.1 & 15.2$\pm$0.2 & (4.3$\pm$0.9)$\times$10$^7$ &  3.2$\pm$1.2 \\
2012 Feb 13 & Gemini-N &  2 &   360 & $r'$  &    56.5 & 2.635 & 2.687 & 21.3 & 20.82$\pm$0.05 & 20.5$\pm$0.1 & 15.2$\pm$0.2 & (4.2$\pm$0.9)$\times$10$^7$ &  5.8$\pm$1.5 \\
2012 Feb 15 & Gemini-N &  9 &  1620 & $r'$  &    57.0 & 2.639 & 2.716 & 21.2 & 20.90$\pm$0.05 & 20.4$\pm$0.1 & 15.1$\pm$0.2 & (4.9$\pm$1.0)$\times$10$^7$ &  5.4$\pm$1.4 \\
2012 Feb 21 & Gemini-N &  2 &   360 & $r'$  &    58.5 & 2.649 & 2.802 & 20.7 & 21.03$\pm$0.05 & 20.8$\pm$0.1 & 15.5$\pm$0.2 & (3.2$\pm$0.7)$\times$10$^7$ &  4.6$\pm$1.3 \\
2016 Jun  8 & CFHT     &  4 &   720 & $r'$  & $-$42.1 & 2.546 & 2.294 & 23.5 & 21.36$\pm$0.05 & 21.4$\pm$0.1 & 16.5$\pm$0.2 & (0.4$\pm$0.3)$\times$10$^7$ &  1.4$\pm$0.1 \\ 
2016 Jul  8 & CFHT     &  9 &  1620 & $r'$  & $-$34.2 & 2.509 & 1.905 & 21.6 & 21.00$\pm$0.05 & 21.0$\pm$0.1 & 16.6$\pm$0.2 & (0.3$\pm$0.3)$\times$10$^7$ &  1.0$\pm$0.1 \\ 
2016 Sep  6 & CFHT     &  5 &   900 & $r'$  & $-$17.8 & 2.456 & 1.452 &  2.8 & 19.15$\pm$0.05 & 19.0$\pm$0.1 & 16.0$\pm$0.1 & (1.6$\pm$0.3)$\times$10$^7$ &  5.4$\pm$1.3 \\ 
2016 Sep  8 & CFHT     &  5 &   900 & $r'$  & $-$17.2 & 2.454 & 1.450 &  2.4 & 19.09$\pm$0.05 & 19.0$\pm$0.1 & 16.0$\pm$0.1 & (1.5$\pm$0.3)$\times$10$^7$ &  5.7$\pm$1.3 \\ 
2016 Sep 25 & Magellan &  3 &  1050 & $r'$  & $-$12.5 & 2.446 & 1.474 &  7.5 & 19.11$\pm$0.05 & 19.1$\pm$0.1 & 15.8$\pm$0.1 & (2.0$\pm$0.4)$\times$10$^7$ &  8.9$\pm$1.8 \\ 
2016 Sep 25 & CFHT     &  5 &   900 & $r'$  & $-$12.4 & 2.446 & 1.475 &  7.6 & 19.08$\pm$0.05 & 19.0$\pm$0.1 & 15.7$\pm$0.1 & (2.4$\pm$0.5)$\times$10$^7$ &  9.4$\pm$1.9 \\ 
2016 Oct  3 & LOT      &  5 &  1500 & $r'$  & $-$10.1 & 2.442 & 1.512 & 11.0 & 19.10$\pm$0.05 & 19.0$\pm$0.1 & 15.5$\pm$0.1 & (3.1$\pm$0.6)$\times$10$^7$ & 11.7$\pm$2.3 \\ 
2016 Oct  8 & Gemini-S &  5 &   750 & $r'$  &  $-$8.8 & 2.441 & 1.541 & 12.8 & 19.19$\pm$0.05 & 19.1$\pm$0.1 & 15.5$\pm$0.1 & (3.1$\pm$0.6)$\times$10$^7$ & 11.7$\pm$2.4 \\ 
2016 Oct 25 & LOT      &  5 &  1500 & $r'$  &  $-$3.9 & 2.437 & 1.680 & 18.3 & 19.53$\pm$0.05 & 19.3$\pm$0.1 & 15.3$\pm$0.2 & (3.8$\pm$0.8)$\times$10$^7$ & 11.1$\pm$2.5 \\ 
2016 Nov  2 & Magellan &  2 &   400 & $r'$  &  $-$1.7 & 2.436 & 1.756 & 20.1 & 19.51$\pm$0.05 & 19.2$\pm$0.1 & 15.1$\pm$0.2 & (5.2$\pm$1.0)$\times$10$^7$ & 13.3$\pm$2.9 \\ 
2016 Nov  5 & Gemini-S &  5 &   900 & $r'$  &  $-$0.9 & 2.436 & 1.788 & 20.7 & 19.68$\pm$0.05 & 19.3$\pm$0.1 & 15.1$\pm$0.2 & (4.9$\pm$1.0)$\times$10$^7$ & 11.3$\pm$2.6 \\ 
2016 Nov  8 & \multicolumn{4}{l}{\it Perihelion .......................} & 0.0 & 2.436 & 1.823 & 21.3  & ---  & --- & --- & \multicolumn{1}{c}{---} & \multicolumn{1}{c}{---} \\
2016 Nov 28 & Gemini-S &  5 &   900 & $r'$  &     5.6 & 2.438 & 2.059 & 23.5 & 19.87$\pm$0.05 & 19.4$\pm$0.1 & 14.8$\pm$0.2 & (6.8$\pm$1.4)$\times$10$^7$ & 12.7$\pm$2.8 \\
2016 Nov 29 & Gemini-N &  5 &   900 & $r'$  &     6.0 & 2.438 & 2.074 & 23.5 & 19.85$\pm$0.05 & 19.4$\pm$0.1 & 14.8$\pm$0.2 & (6.9$\pm$1.4)$\times$10$^7$ & 13.2$\pm$2.9 \\
2016 Dec 26 & CFHT     &  8 &  1440 & $r'$  &    13.6 & 2.448 & 2.421 & 23.3 & 20.10$\pm$0.05 & 19.6$\pm$0.1 & 14.7$\pm$0.2 & (8.1$\pm$1.6)$\times$10$^7$ & 12.5$\pm$2.7 \\
2016 Dec 29 & CFHT     &  5 &   900 & $r'$  &    14.4 & 2.449 & 2.459 & 23.1 & 20.04$\pm$0.05 & 19.6$\pm$0.1 & 14.6$\pm$0.2 & (8.3$\pm$1.6)$\times$10$^7$ & 13.7$\pm$2.9 \\
2016 Dec 29 & Gemini-N &  4 &   720 & $r'$  &    14.4 & 2.449 & 2.459 & 23.1 & 20.10$\pm$0.05 & 19.7$\pm$0.1 & 14.7$\pm$0.2 & (7.5$\pm$1.5)$\times$10$^7$ & 12.8$\pm$2.7 \\
2017 Jan 15 & Gemini-N &  6 &  1080 & $r'$  &    19.2 & 2.459 & 2.673 & 21.6 & 20.05$\pm$0.05 & 19.7$\pm$0.1 & 14.6$\pm$0.2 & (8.7$\pm$1.6)$\times$10$^7$ & 14.7$\pm$2.9 \\
2022 Mar  2 & \multicolumn{4}{l}{\it Perihelion .......................} & 0.0 & 2.437 & 3.374 &  6.5 & ---  & --- & --- & \multicolumn{1}{c}{---} & \multicolumn{1}{c}{---} \\
\hline
\hline
\multicolumn{14}{l}{$^a$ Telescope used.} \\
\multicolumn{14}{l}{$^b$ Number of exposures.} \\
\multicolumn{14}{l}{$^c$ Total integration time, in seconds.} \\
\multicolumn{14}{l}{$^d$ True anomaly, in degrees.} \\
\multicolumn{14}{l}{$^e$ Heliocentric distance, in au.} \\
\multicolumn{14}{l}{$^f$ Geocentric distance, in au.} \\
\multicolumn{14}{l}{$^g$ Solar phase angle (Sun-object-Earth), in degrees.} \\
\multicolumn{14}{l}{$^h$ Equivalent mean apparent $R$-band nucleus magnitude, measured within photometry apertures with radii of $4\farcs0$.} \\
\multicolumn{14}{l}{$^i$ Equivalent total apparent $R$-band magnitude, including the entire coma and tail, if present.} \\
\multicolumn{14}{l}{$^j$ Total absolute $R$-band magnitude, using $H,G$ phase function where $G=0.18$.} \\
\multicolumn{14}{l}{$^k$ Estimated total dust mass, in kg, assuming $\rho_d\sim2500$~kg~m$^3$.} \\
\multicolumn{14}{l}{$^l$ $Af\rho$ values computed using photometry apertures with radii of $4\farcs0$, in cm, where uncertainties are estimated to be $\sim$10\%.} \\
\end{tabular}
\label{table:obs_288p_active}
\end{table*}

\setlength{\tabcolsep}{4.5pt}
\setlength{\extrarowheight}{0em}
\begin{table*}[htb!]
\caption{Observation Log: 288P - Inactive}
\centering
\smallskip
\footnotesize
\begin{tabular}{lcrrcrccrcc}
\hline\hline
\multicolumn{1}{c}{UT Date}
 & \multicolumn{1}{c}{Tel.$^a$}
 & \multicolumn{1}{c}{$N$$^b$}
 & \multicolumn{1}{c}{$t$$^c$}
 & \multicolumn{1}{c}{Filter}
 & \multicolumn{1}{c}{$\nu$$^d$}
 & \multicolumn{1}{c}{$R$$^e$}
 & \multicolumn{1}{c}{$\Delta$$^f$}
 & \multicolumn{1}{c}{$\alpha$$^g$}
 & \multicolumn{1}{c}{$m_{R,n}$$^h$}
 & \multicolumn{1}{c}{$m_{R,mid}(1,1,\alpha)$$^i$}
 \\
\hline
2012 Oct 18 & Gemini-N & 10 &  1800 & $r'$  & 108.0 & 3.118 & 3.226 & 18.0 & 22.42$\pm$0.05 & 17.39$\pm$0.30 \\
2012 Nov  9 & Gemini-N & 13 &  2340 & $r'$  & 111.7 & 3.160 & 2.953 & 18.2 & 22.90$\pm$0.05 & 18.05$\pm$0.35 \\
2012 Nov 10 & Gemini-N & 10 &  1800 & $r'$  & 111.9 & 3.162 & 2.940 & 18.2 & 22.73$\pm$0.05 & 17.91$\pm$0.35 \\
2012 Nov 13 & Gemini-N & 24 &  4320 & $r'$  & 112.4 & 3.168 & 2.902 & 18.1 & 22.66$\pm$0.05 & 17.68$\pm$0.05 \\
2012 Nov 14 & Gemini-N &  7 &  1260 & $r'$  & 112.6 & 3.170 & 2.890 & 18.0 & 22.85$\pm$0.05 & 18.09$\pm$0.30 \\
2012 Nov 15 & Gemini-N & 10 &  1800 & $r'$  & 112.7 & 3.172 & 2.877 & 18.0 & 22.72$\pm$0.05 & 17.90$\pm$0.30 \\
2012 Nov 20 & Gemini-N &  3 &   540 & $r'$  & 113.6 & 3.181 & 2.815 & 17.6 & 22.80$\pm$0.05 & 18.04$\pm$0.40 \\
2012 Dec  5 & Gemini-N & 14 &  2520 & $r'$  & 116.1 & 3.208 & 2.637 & 15.9 & 22.51$\pm$0.05 & 17.86$\pm$0.40 \\
2012 Dec 14 & Gemini-N & 35 &  6300 & $r'$  & 117.5 & 3.225 & 2.541 & 14.2 & 21.77$\pm$0.05 & 17.23$\pm$0.35 \\
2012 Dec 18 & UH2.2    & 12 &  7200 & $R$   & 118.2 & 3.232 & 2.503 & 13.4 & 22.21$\pm$0.05 & 17.66$\pm$0.40 \\
2012 Dec 19 & UH2.2    & 10 &  6000 & $R$   & 118.3 & 3.234 & 2.494 & 13.1 & 21.85$\pm$0.05 & 17.37$\pm$0.40 \\
2013 Jan  4 & CFHT     &  5 &   750 & $r'$  & 120.9 & 3.262 & 2.374 &  8.7 & 21.66$\pm$0.05 & 17.36$\pm$0.40 \\
2013 Jan 16 & CFHT     &  6 &   900 & $r'$  & 122.8 & 3.282 & 2.325 &  4.7 & 21.48$\pm$0.05 & 17.09$\pm$0.40 \\
2013 Jan 17 & CFHT     &  6 &   900 & $r'$  & 123.0 & 3.284 & 2.323 &  4.4 & 21.64$\pm$0.05 & 17.24$\pm$0.15 \\
2013 Jan 18 & CFHT     &  1 &   150 & $r'$  & 123.1 & 3.286 & 2.321 &  4.0 & 21.55$\pm$0.05 & 17.14$\pm$0.40 \\
2013 May 12 & UH2.2    &  8 &  4800 & $R$   & 140.0 & 3.457 & 3.551 & 16.5 & 23.2$\pm$0.1   & 17.8$\pm$0.4   \\
2013 May 13 & UH2.2    &  7 &  4200 & $R$   & 140.2 & 3.458 & 3.566 & 16.4 & 23.0$\pm$0.1   & 17.5$\pm$0.4   \\
2015 Apr 24 & CFHT     &  2 &   360 & $r'$  & $-$125.0 & 3.307 & 2.438 & 10.3 & 21.61$\pm$0.05 & 17.07$\pm$0.40 \\ 
2015 May 26 & CFHT     &  2 &   360 & $r'$  & $-$120.0 & 3.252 & 2.239 &  0.5 & 21.25$\pm$0.05 & 16.94$\pm$0.40 \\ 
2015 May 27 & CFHT     &  6 &   900 & $r'$  & $-$119.9 & 3.250 & 2.237 &  0.7 & 21.13$\pm$0.05 & 16.84$\pm$0.35 \\ 
\hline
\hline
\multicolumn{11}{l}{$^a$ Telescope used (UH2.2: UH 2.2~m telescope).} \\
\multicolumn{11}{l}{$^b$ Number of exposures.} \\
\multicolumn{11}{l}{$^c$ Total integration time, in seconds.} \\
\multicolumn{11}{l}{$^d$ True anomaly, in degrees.} \\
\multicolumn{11}{l}{$^e$ Heliocentric distance, in au.} \\
\multicolumn{11}{l}{$^f$ Geocentric distance, in au.} \\
\multicolumn{11}{l}{$^g$ Solar phase angle (Sun-object-Earth), in degrees.} \\
\multicolumn{11}{l}{$^h$ Equivalent mean apparent $R$-band nucleus magnitude.} \\
\multicolumn{11}{l}{$^i$ Estimated reduced $R$-band magnitude at midpoint of full photometric range (assumed to be 0.80 mag)} \\
\multicolumn{11}{l}{$~~~$ of rotational light curve.} \\
\end{tabular}
\label{table:obs_288p_inactive}
\end{table*}

\setlength{\tabcolsep}{5pt}
\begin{table*}[htb!]
\caption{Previously reported observations of 288P when active}
\centering
\smallskip
\footnotesize
\begin{tabular}{lcrccrccccc}
\hline\hline
\multicolumn{1}{c}{UT Date}
 & \multicolumn{1}{c}{Tel.$^a$}
 & \multicolumn{1}{c}{$\nu$$^b$}
 & \multicolumn{1}{c}{$R$$^c$}
 & \multicolumn{1}{c}{$\Delta$$^d$}
 & \multicolumn{1}{c}{$\alpha$$^e$}
 & \multicolumn{1}{c}{$m_{R,n}$$^f$}
 & \multicolumn{1}{c}{$m_{R,t}$$^g$}
 & \multicolumn{1}{c}{$H_{R,t}$$^h$}
 & \multicolumn{1}{c}{$M_d$$^i$}
 & \multicolumn{1}{c}{Ref.$^j$}
 \\
\hline
2011 Jul 18 & \multicolumn{1}{l}{\it Perihelion ...} &   0.0 & 2.438 & 2.292 & 24.6 & --- & --- & --- & \multicolumn{1}{c}{---} & --- \\
2011 Aug 30 & PS1       &  12.2 & 2.447 & 1.806 & 21.4 & 20.16$\pm$0.14 & $<$20.2 & $<$15.9 & $>$(1.6$\pm$0.8)$\times$10$^7$ & [1] \\
2011 Nov  5 & PS1       &  30.7 & 2.496 & 1.517 &  4.6 & 19.00$\pm$0.09 & $<$19.0 & $<$15.7 & $>$(2.2$\pm$1.0)$\times$10$^7$ & [1] \\
2011 Nov 12 & FTN       &  32.9 & 2.505 & 1.555 &  8.0 & 18.69$\pm$0.09 & $<$18.7 & $<$15.2 & $>$(4.5$\pm$1.6)$\times$10$^7$ & [1] \\
2011 Nov 14 & UH2.2     &  33.2 & 2.506 & 1.561 &  8.4 & 18.62$\pm$0.05 & $<$18.6 & $<$15.1 & $>$(5.0$\pm$1.8)$\times$10$^7$ & [1] \\
2011 Nov 14 & FTN       &  33.2 & 2.506 & 1.561 &  8.4 & 18.64$\pm$0.05 & $<$18.6 & $<$15.1 & $>$(4.9$\pm$1.8)$\times$10$^7$ & [1] \\
2011 Nov 18 & Perkins   &  34.3 & 2.510 & 1.586 & 10.0 & 18.60$\pm$0.10 & $<$18.6 & $<$15.0 & $>$(5.8$\pm$2.0)$\times$10$^7$ & [1] \\
2011 Nov 19 & HCT       &  34.6 & 2.512 & 1.596 & 10.6 & 18.64$\pm$0.02 & $<$18.6 & $<$15.0 & $>$(5.8$\pm$2.0)$\times$10$^7$ & [1] \\
2011 Nov 22 & WHT       &  35.5 & 2.516 & 1.621 & 11.9 & 18.89$\pm$0.02 & $<$18.9 & $<$15.1 & $>$(4.8$\pm$1.7)$\times$10$^7$ & [1] \\
2011 Nov 30 & UH2.2     &  37.4 & 2.525 & 1.685 & 14.4 & 19.04$\pm$0.02 & $<$19.0 & $<$15.1 & $>$(4.9$\pm$1.8)$\times$10$^7$ & [1] \\
2011 Dec  4 & NTT       &  38.5 & 2.530 & 1.724 & 15.6 & 19.12$\pm$0.03 & $<$19.1 & $<$15.1 & $>$(5.0$\pm$1.8)$\times$10$^7$ & [1] \\
2011 Dec 16 & FTS       &  41.7 & 2.546 & 1.861 & 18.7 & 19.70$\pm$0.09 & $<$19.7 & $<$15.4 & $>$(3.5$\pm$1.4)$\times$10$^7$ & [1] \\
2011 Dec 19 & UH2.2     &  42.4 & 2.549 & 1.895 & 19.2 & 19.68$\pm$0.03 & $<$19.7 & $<$15.3 & $>$(3.8$\pm$1.5)$\times$10$^7$ & [1] \\
2012 Jan  7 & LOT       &  47.4 & 2.577 & 2.152 & 21.7 & 20.43$\pm$0.10 & $<$20.4 & $<$15.7 & $>$(2.3$\pm$1.0)$\times$10$^7$ & [1] \\
2016 Nov  8 & \multicolumn{1}{l}{\it Perihelion ...} &   0.0 & 2.436 & 1.823 & 21.3 & --- & --- & --- & \multicolumn{1}{c}{---} & --- \\
\hline
\hline
\multicolumn{11}{l}{$^a$ Telescope (PS1: Pan-STARRS1; FTN: Faulkes Telescope North; UH2.2: UH 2.2~m telescope; Perkins: Lowell } \\
\multicolumn{11}{l}{$~~~~$Observatory Perkins Telescope; HCT: Himalayan Chandra Telescope; WHT: William Herschel Telescope; } \\
\multicolumn{11}{l}{$~~~~$NTT: New Technology Telescope; FTS: Faulkes Telescope South; LOT: Lulin One-meter Telescope).} \\
\multicolumn{11}{l}{$^b$ True anomaly, in degrees.} \\
\multicolumn{11}{l}{$^c$ Heliocentric distance, in au.} \\
\multicolumn{11}{l}{$^d$ Geocentric distance, in au.} \\
\multicolumn{11}{l}{$^e$ Solar phase angle (Sun-object-Earth), in degrees.} \\
\multicolumn{11}{l}{$^f$ Reported mean apparent $R$-band nucleus magnitude.} \\
\multicolumn{11}{l}{$^g$ Equivalent mean apparent $R$-band total magnitude, including the entire coma and tail, if present.} \\
\multicolumn{11}{l}{$^h$ Absolute $R$-band total magnitude (at $R$$\,=\,$$\Delta$$\,=\,$1~au and $\alpha$$\,=\,$0$^{\circ}$),} \\
\multicolumn{11}{l}{$~~~~$using IAU $H,G$ phase-darkening where $G=0.18$.} \\
\multicolumn{11}{l}{$^i$ Estimated total dust mass, in kg, assuming $\rho_d\sim2500$~kg~m$^3$.} \\
\multicolumn{11}{l}{$^j$ [1] \citet{hsieh2012_288p}} \\
\\
\\
\end{tabular}
\label{table:obs_288p_active_prev}
\end{table*}

Observations of 238P and 288P presented here were obtained with the 8.1~m Gemini North (Gemini-N) telescope, the 3.54~m Canada-France-Hawaii Telescope (CFHT), and the University of Hawaii (UH) 2.2~m telescope on Maunakea in Hawaii, the 8.1~m Gemini South (Gemini-S) telescope at Cerro Pachon in Chile, the 6.5~m Baade Magellan telescope at Las Campanas in Chile, Lowell Observatory's 4.3~m Discovery Channel Telescope (DCT) at Happy Jack, Arizona, the 2.5~m Sloan Digital Sky Survey (SDSS) telescope at Apache Point Observatory in New Mexico, and the Lulin One-meter Telescope (LOT) at Lulin Observatory in Taiwan.  We employed the Gemini Multi-Object Spectrographs \citep[GMOS;][]{hook2004_gmos,gimeno2016_gmoss} and Sloan $r'$-band filters for Gemini-N and Gemini-S observations, MegaCam \citep{boulade2003_megacam} and a Sloan $r'$-band filter for CFHT observations, a 2048$\times$2048 pixel Textronix CCD and a Kron-Cousins $R$-band filter for UH 2.2~m observations, the Inamori Magellan Areal Camera and Spectrograph \citep[IMACS;][]{dressler2011_imacs} and a Sloan $r'$-band filter for Baade observations, the Large Monolithic Imager \citep{bida2014_dct} and a Kron-Cousins $R$-band filter for DCT observations, and a VersArray:1300B CCD \citep{kinoshita2005_lulin} and a Bessell-like $R$-band filter for LOT observations. SDSS data presented here were obtained using a large-format mosaic CCD camera designed for the SDSS survey \citep{gunn1998_sdss} and a Sloan $r'$-band filter.  Non-sidereal tracking was used for all targeted observations, while sidereal tracking was used for SDSS survey observations.

Observations of 238P while it was active in 2011 were obtained using Gemini-N (Program GN-2011B-Q-17), and represent a continuation of the observing campaign previously described in \citet{hsieh2011_238p}.  Observations of 238P over several months while it was most recently active in 2016 and 2017 were obtained by Gemini-N (Program GN-2016B-LP-11), CFHT (Program 16BT05), LOT, and DCT.  Details of these observations are listed in Table~\ref{table:obs_238p_active}.  Orbit positions of both the observations reported here and previously reported observations \citep[Table~\ref{table:obs_238p_active_prev};][]{hsieh2009_238p,hsieh2011_238p} are marked in Figure~\ref{figure:orbit_238p}.  Composite images of the object during each night of newly reported observations are shown in Figure~\ref{figure:images_238p_active}.

Meanwhile, observations of 288P while it was active in 2012 were obtained by Gemini-N (Program GN-2012A-Q-68), and represent a continuation of the observing campaign previously described in \citet{hsieh2012_288p}.  Since then, we also obtained observations of 288P while it was inactive in 2012 with Gemini-N (Program GN-2012B-Q-106), in 2012 and 2013 with the UH 2.2~m telescope, and in 2013 and 2015 with CFHT (Programs 12BH43 and 15AT05) as part of a campaign to characterize its nucleus, and also obtained observations of the object while it was most recently active in 2016 and 2017 using Gemini-N and Gemini-S (Programs GN-2016B-LP-11 and GS-2016B-LP-11), CFHT (Program 16BT05), Magellan, and LOT.  In addition, using the Solar System Object Image Search tool\footnote{\tt http://www.cadc-ccda.hia-iha.nrc-cnrc.gc.ca/en/ssois/} \citep[SSOIS;][]{gwyn2012_ssois}, provided by the Canadian Astronomical Data Centre, we also identified precovery observations obtained in 2000 by the SDSS survey \citep{abazajian2009_sdssdr7,york2000_sdss,fukugita1996_sdss,gunn1998_sdss,gunn2006_sdss,aihara2011_sdss} in which the object appeared to be active.  Details of these observations of 288P are listed in Tables~\ref{table:obs_288p_active} and \ref{table:obs_288p_inactive}.  Orbit positions of both the observations reported here and previously reported observations \citep[Table~\ref{table:obs_288p_active_prev};][]{hsieh2012_288p} are marked in Figure~\ref{figure:orbit_288p}.  Composite images of the object during each night of observations when it was active are shown in Figure~\ref{figure:images_288p_active}, while composite images of the object during each night of observations when it was inactive are shown in Figure~\ref{figure:images_288p_inactive}.

We performed standard bias subtraction and flat-field reduction (using dithered images of the twilight sky) for all data from targeted observations, except those from CFHT, using IRAF software \citep{tody1986_iraf,tody1993_iraf}.  Reduction of CFHT data was performed by the Elixir pipeline \citep{magnier2004_elixir}. Photometric calibration of UH 2.2~m data was accomplished using \citet{landolt1992_standardstars} standard stars and field stars, for which net fluxes were measured within circular apertures, with background sampled from surrounding circular annuli.  For Gemini-N, Gemini-S, CFHT, Magellan, DCT, and LOT data, absolute photometric calibration was accomplished using field star magnitudes from SDSS or Pan-STARRS1 field star catalogs \citep{aihara2011_sdss,schlafly2012_ps1,tonry2012_ps1,magnier2013_ps1,magnier2016_ps1photometryastrometry,flewelling2016_ps1}. Conversion of $r'$-band Gemini and PS1 photometry to $R$-band was accomplished using transformations derived by \citet{tonry2012_ps1} and by R.\ Lupton ({\tt http://www.sdss.org/}). Comet photometry was performed using circular apertures with $4\farcs0$ radii, where background statistics were measured in nearby, but non-adjacent, regions of blank sky to avoid dust contamination from the comet. To maximize signal-to-noise ratios of cometary features, we constructed composite images of the object for each night of data by shifting and aligning individual images on the object's photocenter using linear interpolation and then adding them together.

In addition to performing nucleus photometry, we also measured the total flux from each object in our composite images from each night when they were observed to be active.  We did so by using rectangular photometry apertures enclosing the entire visible dust cloud and oriented to avoid field star contamination. Background sky levels were then measured from nearby areas of blank sky and subtracted to obtain net fluxes.  These net fluxes were then calibrated using standard stars or field stars from SDSS or PS1 to obtain absolute photometry.

\section{RESULTS AND ANALYSIS}\label{section:results}

\subsection{288P Phase Function Analysis}\label{section:288p_inactive}

To enable detailed analysis of 288P's activity, we first need to characterize the properties of its nucleus, specifically its phase function, from which we can derive its size and expected brightnesses at given viewing geometries \citep[where 238P's nucleus phase function has already been characterized by][]{hsieh2011_238p}.  In order to determine the phase function of 288P's nucleus, we only consider photometric data obtained for the object when no visible activity is detected in stacked composite images (Figure~\ref{figure:images_288p_inactive}) and the orbit position of the object is also sufficiently far from perihelion that significant activity is not expected, i.e., data from late 2012 to 2015.  We normalize apparent magnitudes, $m(R,\Delta,\alpha)$, meeting these criteria to unit heliocentric and geocentric distances, $R$ and $\Delta$, respectively, where $\alpha$ is the solar phase angle, to obtain reduced magnitudes, $m(1,1,\alpha)$, using
\begin{equation}
m(1,1,\alpha) = m(R,\Delta,\alpha) - 5\log(R\Delta)
\end{equation}
The majority of our observations were short-duration ``snapshot'' observations at unknown rotational phases.  For this analysis, we assume that the sparsely sampled nature of our data means that various deviations of our photometric data from the rotationally averaged brightness of the nucleus (i.e., the ``mid-point'' of its rotational lightcurve) at the times of our observations average to zero, allowing us to fit a phase function solution to our data that reflects that rotationally averaged nucleus brightness.  To account for nights when at least partial lightcurves were obtained (i.e., where some photometric variation is clearly present), we compute the uncertainty, $\sigma_m$, for the average magnitude of each night's observations using
\begin{equation}
\sigma_{m} = {\Delta m_{\rm exp} - \Delta m_{\rm obs}\over 2}
\end{equation}
where $\Delta m_{\rm exp}$ is the expected or assumed total photometric range and $\Delta m_{\rm obs}$ is the observed photometric range. \citet{waniak2016_288p} have measured the lightcurve of 288P to have an amplitude (i.e., peak to midpoint) of 0.4~mag, and as such, we use $\Delta m_{\rm exp}=0.8$~mag as the expected total photometric range for this analysis.

We use these magnitude uncertainties to compute weighted average magnitudes ($m_{R,mid}(1,1,\alpha)$ in Table~\ref{table:obs_288p_inactive}) over the time periods in which we are interested (where in all cases, this assumed rotational uncertainty dominates the photometric uncertainties for all of our data), and then perform a least-squares fit to this data to obtain the best-fit phase function.

We find best-fit phase function parameters for 288P's nucleus of $H_R=16.80\pm0.12$~mag and $G_R=0.18\pm0.11$.  Assuming solar colors (i.e., $V-R=0.36$), these results correspond to a $V$-band absolute magnitude of $H_V\sim17.16\pm0.12$~mag, consistent with the results of \citet{agarwal2016_288p} (who assumed $G=0.15$ and did not perform a full phase function fit) within uncertainties.  Assuming a $R$-band albedo of $p_R$$\,=\,$0.05, this absolute magnitude corresponds to an effective nucleus radius of $r_N=1.13\pm0.06$~km or, assuming 288P's nucleus to be a binary system with approximately equally sized components \citep{agarwal2017_288p}, effective component radii of $r_c=0.80\pm0.04$~km each.

\begin{figure}[htb!]
\centerline{\includegraphics[width=3.3in]{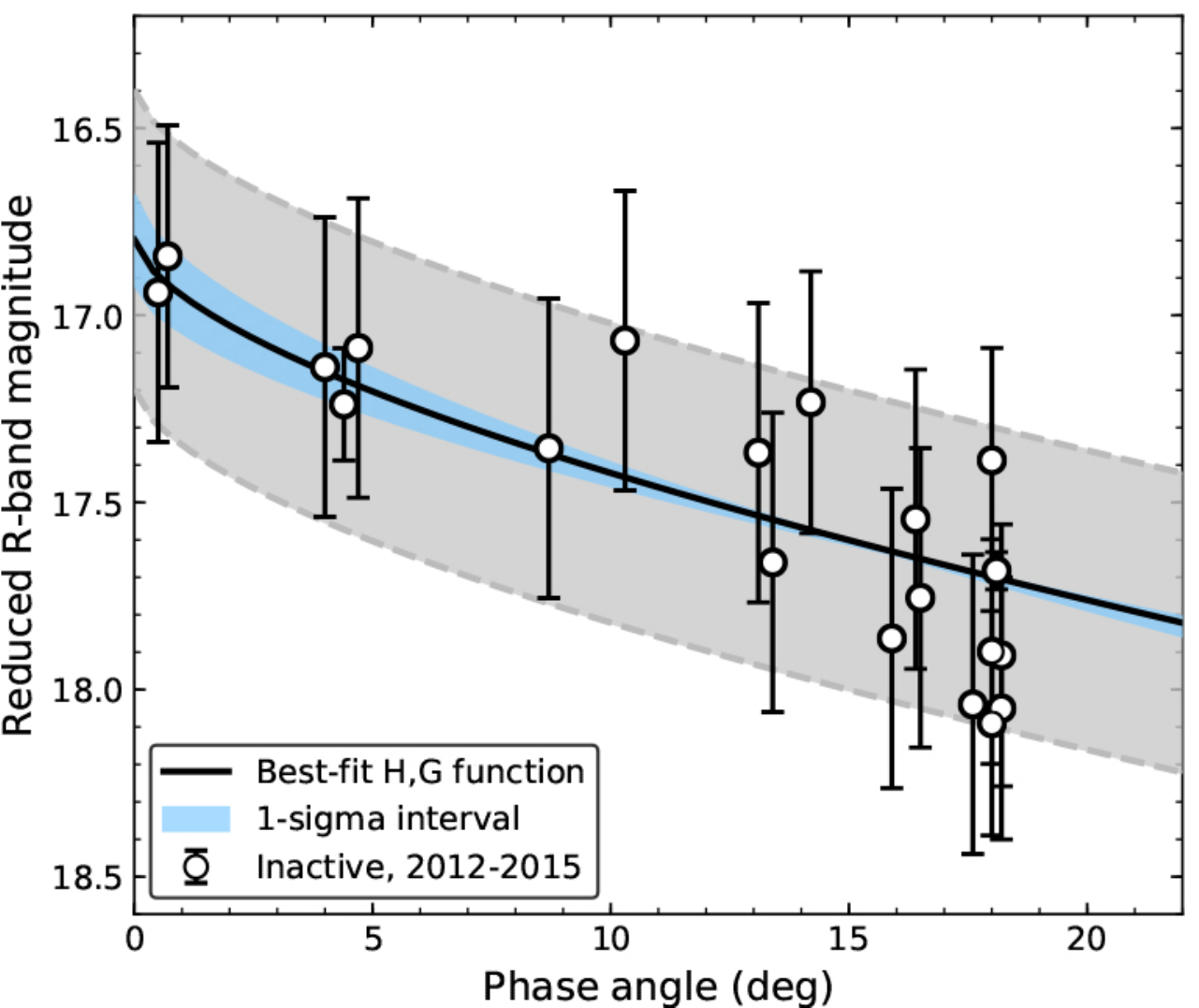}}
\caption{\small Best-fit IAU phase function (solid line) for 288P where estimated reduced $R$-band magnitudes at the midpoint of the full photometric range (assumed to be 0.8 mag) of the object's rotational light curve for data obtained between 2012 October to 2015 May when no activity was detected for 288P are plotted with open circles.  Dotted lines indicate the range of uncertainty due to estimated phase function parameter uncertainties, while dashed lines indicate the possible photometric range due to rotational brightness variations, assuming a peak-to-trough photometric range of $\Delta m=0.8$~mag. 
}
\label{figure:phase_function_288p}
\end{figure}

\subsection{Overview of Photometric Activity Characterization}\label{section:photometric_analysis}

We compute absolute total $R$-band magnitudes, $H_{R,t}$, for 238P and 288P (i.e., at $R=\Delta=1$~au and $\alpha=0^{\circ}$) from measured apparent total magnitudes, $m_{R,t}$, for each set of observations (Tables~\ref{table:obs_238p_active}, \ref{table:obs_288p_inactive}, and \ref{table:obs_288p_active}), assuming inverse-square law fading and $H,G$ phase functions with $G$$\,=\,$$-0.03$ for 238P \citep{hsieh2011_238p} and $G$$\,=\,$0.18 for 288P (Section~\ref{section:288p_inactive}) as computed for each object's nucleus.  Emitted dust may not have the same photometric behavior as the nucleus though, making these assumptions a source of uncertainty.  From the observed photometric excesses above the expected brightness of each object's inactive nucleus \citep[using $H_R$$\,=\,$19.05~mag for 238P and $H_R$$\,=\,$16.80~mag for 288P; Section~\ref{section:288p_inactive} of this work;][]{hsieh2011_238p}, we then estimate the total mass, $M_d$, of visible ejected dust using
\begin{equation}
M_d = {4\over 3}\pi r^2_N {\bar a}\rho_d \left({1-10^{0.4(H_{R,t}-H_R)} \over 10^{0.4(H_{R,t}-H_R)} }\right)
\end{equation}
\citep[cf.][]{hsieh2014_324p}, where $r_N$$\,=\,$0.4~km is the estimated effective nucleus radius for 238P \citep{hsieh2011_238p} and $r_c$$\,=\,$0.8~km is the estimated effective radius for each component of 288P's binary nucleus (Section~\ref{section:288p_inactive}), assuming approximately equally sized components \citep[cf.][]{agarwal2017_288p}.

We assume dust grain densities of $\rho_d$$\,=\,$2500~kg~m$^{-3}$, consistent with CI and CM carbonaceous chondrites, which are associated with primitive C-type objects like the MBCs \citep{britt2002_astdensities_ast3}, and effective mean dust grain radii of ${\bar a}$$\,=\,$1~mm \citep[assuming power-law particle size distributions from $\mu$m- to cm-sized particles determined from dust modeling of other MBCs;][]{moreno2011_324p,hsieh2014_324p}.  For reference, we also compute and report $Af\rho$ values \citep{ahearn1984_bowell}, although we note that this parameter is not always a reliable measurement of the dust contribution to comet photometry in cases of non-spherically symmetric comae \citep[e.g.,][]{fink2012_afrho}.  The results of these calculations are shown in Tables~\ref{table:obs_238p_active} and \ref{table:obs_288p_active}.

\subsection{238P Activity Characterization}\label{section:238p_active}

\setlength{\tabcolsep}{4.0pt}
\setlength{\extrarowheight}{0em}
\begin{table*}[htb!]
\caption{Estimated Activity Start Times and Strength Parameters}
\centering
\smallskip
\footnotesize
\begin{tabular}{ccccccc}
\hline\hline
\multicolumn{1}{c}{\rule{0pt}{2.4ex} Object}
 & \multicolumn{1}{c}{Data range$^a$}
 & \multicolumn{1}{c}{Fit results$^b$}
 & \multicolumn{1}{c}{Midpoint$^c$}
 & \multicolumn{1}{c}{${\dot m}_{\rm H_2O}$$^d$}
 & \multicolumn{1}{c}{$A_{\rm act}$$^e$}
 & \multicolumn{1}{c}{$f_{\rm act}$$^f$}
 \\
\hline
\rule{0pt}{2.6ex}238P & 2010 Sep 3 $-$ 2010 Dec 9                           & ${\dot M}_d=1.4\pm0.3$~kg~s$^{-1}$  & 2010 Oct 21       & $2.1\times10^{-6}$ (IT) & $1\times10^5$ (IT) & $7\times10^{-2}$ (IT) \\
\rule{0pt}{2.5ex}     & $R$: $2.576~{\rm au} \rightarrow 2.416~{\rm au}$    & $t_0 = 2010~{\rm Aug}~17\pm50$~days & $R=2.49$~au       & $5.0\times10^{-5}$ (SS) & $6\times10^3$ (SS) & $3\times10^{-3}$ (SS) \\
\rule{0pt}{2.5ex}     & $\nu$: $-54.2^{\circ} \rightarrow -27.6^{\circ}$    & $R_0 = 2.61\mp0.11$~au              & $\nu=-41^{\circ}$ \\
\rule{0pt}{2.5ex}     & \vspace{0.03cm}                                     & $\nu_0 = -58\pm12^{\circ}$             \\ 
\hline
\rule{0pt}{2.6ex}238P & 2016 Jul 8 $-$ 2016 Oct 9                           & ${\dot M}_d=0.7\pm0.3$~kg~s$^{-1}$  & 2016 Aug 24       & $3.0\times10^{-6}$ (IT) & $5\times10^4$ (IT) & $2\times10^{-2}$ (IT) \\
\rule{0pt}{2.5ex}     & $R$: $2.439~{\rm au} \rightarrow 2.367~{\rm au}$    & $t_0 = 2016~{\rm Mar}~11\pm85$~days & $R=2.39$~au       & $5.6\times10^{-5}$ (SS) & $3\times10^3$ (SS) & $1\times10^{-3}$ (SS) \\
\rule{0pt}{2.5ex}     & $\nu$: $-31.5^{\circ} \rightarrow -4.0^{\circ}$     & $R_0 = 2.66\mp0.19$~au              & $\nu=-18^{\circ}$ \\
\rule{0pt}{2.5ex}     & \vspace{0.03cm}                                     & $\nu_0 = -63\pm21^{\circ}$          \\ 
\hline
\hline
\rule{0pt}{2.6ex}288P & 2000 Sep 3 $-$ 2000 Nov 17                          & ${\dot M}_d=3.5\pm0.4$~kg~s$^{-1}$  & 2000 Oct 11        & $2.3\times10^{-6}$ (IT) & $3\times10^5$ (IT) & $2\times10^{-2}$ (IT) \\
\rule{0pt}{2.5ex}     & $R$: $2.491~{\rm au} \rightarrow 2.464~{\rm au}$    & $t_0 = 2000~{\rm Aug}~9\pm15$~days  & $R=2.47$~au        & $5.1\times10^{-5}$ (SS) & $1\times10^4$ (SS) & $9\times10^{-4}$ (SS) \\
\rule{0pt}{2.5ex}     & $\nu$: $-21.1^{\circ} \rightarrow -0.4^{\circ}$     & $R_0 = 2.51\mp0.02$~au              & $\nu=-11^{\circ}$  \\
\rule{0pt}{2.5ex}     & \vspace{0.03cm}                                     & $\nu_0 = -28\pm4^{\circ}$           \\ 
\hline
\rule{0pt}{2.6ex}288P & 2016 Sep 6 $-$ 2016 Oct 25                          & ${\dot M}_d=5.6\pm0.7$~kg~s$^{-1}$  & 2016 Sep 30         & $2.5\times10^{-6}$ (IT) & $4\times10^5$ (IT) & $3\times10^{-2}$ (IT) \\
\rule{0pt}{2.5ex}     & $R$: $2.546~{\rm au} \rightarrow 2.437~{\rm au}$    & $t_0 = 2016~{\rm Aug}~5\pm15$~days  & $R=2.44$~au         & $5.3\times10^{-5}$ (SS) & $2\times10^4$ (SS) & $1\times10^{-3}$ (SS) \\
\rule{0pt}{2.5ex}     & $\nu$: $-42.3^{\circ} \rightarrow -4.0^{\circ}$     & $R_0 = 2.48\mp0.02$~au              & $\nu=-11^{\circ}$   \\
\rule{0pt}{2.5ex}     & \vspace{0.03cm}                                     & $\nu_0 = -27\pm4^{\circ}$           \\ 
\hline
\hline
\multicolumn{7}{l}{\rule{0pt}{2.6ex}$^a$ Range of data used for fitting analysis, including dates and corresponding ranges of heliocentric distances, $R$, and true anomalies, $\nu$.} \\
\multicolumn{7}{l}{$^b$ Best-fit results from fitting analysis for initial net dust production rate, ${\dot M}_d$, start date, $t_0$, heliocentric distance at start date, $R_0$,} \\
\multicolumn{7}{l}{$~~~$  and true anomaly at start date, $\nu_0$.} \\
\multicolumn{7}{l}{$^c$ Date at midpoint of range of data used for fitting analysis, including heliocentric distance, $R$, and true anomaly, $\nu$, on that date.} \\
\multicolumn{7}{l}{$^d$ Expected sublimation rates of water for a sublimating graybody in thermal equilibrium, in kg~s$^{-1}$~m$^{-2}$, at the heliocentric distance} \\
\multicolumn{7}{l}{$~~~$    of the object at the indicated midpoint date, using isothermal (IT) and subsolar (SS) approximations.} \\
\multicolumn{7}{l}{$^e$ Estimated effective active areas, in m$^2$, assuming $f_{dg}=5$, implied by best-fit initial net dust production rates and expected} \\
\multicolumn{7}{l}{$~~~$    sublimation rates of water computed using isothermal (IT) and subsolar (SS) approximations.} \\
\multicolumn{7}{l}{$^f$ Estimated effective active fractions, assuming $f_{dg}=5$, and $r_N=400$~m for 238P's nucleus and $r_c=800$~m for each component of} \\
\multicolumn{7}{l}{$~~~$    288P's binary nucleus, for expected sublimation rates of water computed using isothermal (IT) and subsolar (SS) approximations.} \\
\end{tabular}
\label{table:activity_fitting_results}
\end{table*}

\begin{figure*}[htb!]
\centerline{\includegraphics[width=5.5in]{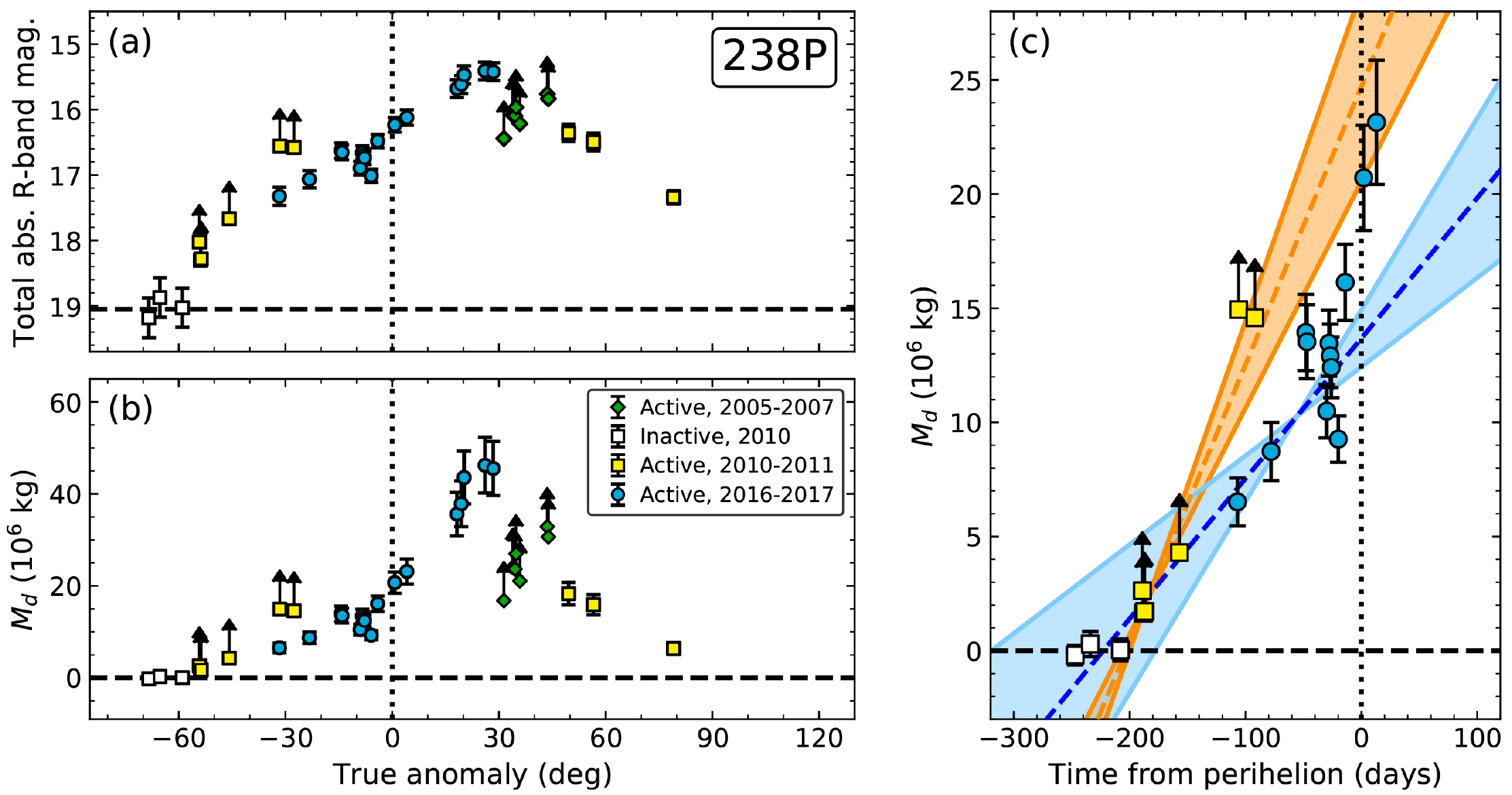}}
\caption{\small (a) Total absolute $R$-band magnitude of 238P during its 2005-2007 active period (green diamonds), 2010 inactive period (open squares), 2010-2011 active period (yellow squares), and 2016-2017 active period (blue circles) plotted as a function of true anomaly.  The expected magnitude of the inactive nucleus is marked with a horizontal dashed black line, while perihelion is marked with a dotted vertical line.  (b) Total estimated dust masses measured for 238P during the same periods of observations as in (a) plotted as a function of true anomaly.  The excess dust mass expected for the inactive nucleus (i.e., zero) is marked with a horizontal dashed black line, while perihelion is marked with a dotted vertical line.  (c)  Estimated total dust masses measured for 238P during just its 2016-2017 active period plotted as a function of time from perihelion (where negative values denote time before perihelion and positive values denote time after perihelion).  A diagonal solid blue line shows a linear fit to data obtained between 2016 July 8 and 2016 November 5 ($-$31.5$^{\circ}$$\,<\,$$\nu$$\,<\,$4.2$^{\circ}$), reflecting the average net dust production rate over this period (over which dust production appears to be roughly linear) and allowing us to estimate the onset time of activity, while diagonal dotted blue lines show the range of uncertainty of the linear fit.  A diagonal solid orange line shows a linear fit to data obtained between 2010 September 3 and 2010 December 9 ($-$54.1$^{\circ}$$\,<\,$$\nu$$\,<\,$$-$27.5$^{\circ}$), reflecting the average net dust production rate over this period (over which dust production appears to be roughly linear), while diagonal dotted orange lines show the range of uncertainty of the linear fit.
}
\label{figure:images_238p_activityevolution}
\end{figure*}

To compare 238P's activity in 2016-2017 to previous active epochs, we compute $M_d$ and $Af\rho$ for previously reported observations of 238P from \citet{hsieh2009_238p} and \citet{hsieh2011_238p}.  Total magnitudes of 238P (i.e., including the entire coma and tail) were not measured for these data in the same way as we have measured recent data, and as such, we report the equivalent measured dust masses as lower limits in Table~\ref{table:obs_238p_active_prev}.  We plot total absolute $R$-band magnitudes and equivalent estimated total dust masses for 238P in Figure~\ref{figure:images_238p_activityevolution}.

We fit a linear function to a portion of the data obtained during the object's 2016-2017 active period, aiming to estimate the initial net dust production rate, ${\dot M}_d$, over this period as well as the approximate onset time of the observed activity.  Specifically, we fit data obtained from 2016 July 8 to 2016 October 9 (where equivalent $R$ and $\nu$ ranges are listed in Table~\ref{table:activity_fitting_results}), when both a reasonable number of data points is available for fitting purposes and measured excess dust masses appear to increase approximately linearly.  Following the calculations detailed by \citet[][]{hsieh2015_ps1mbcs}, we find that the heliocentric distance change over this period corresponds to a $\sim$9\%--28\% increase in the water sublimation rate on the object's surface, depending on whether the subsolar or isothermal approximation is assumed.  The resulting dust production rate and corresponding activity start date we find are of course subject to numerous sources of uncertainty including the nonlinearity of the actual dust production rate as a function of heliocentric distance, ordinary photometric calibration uncertainties, uncertainties specifically associated with measuring extended objects (e.g., selection of optimal photometry apertures), and the unknown rotational phases of the object at the times when each photometric point was obtained.

We find a best-fit initial average net dust production rate of ${\dot M}_d$$\,=\,$0.7$\pm$0.3~kg~s$^{-1}$ over the time period specified above, and a best-fit start date for activity of $\sim225\pm85$~days prior to perihelion (corresponding to a best-fit start date of 2016 March 11, and an uncertainty range of 2015 December 17 to 2016 June 4), when the object was at $R=2.66\mp0.19$~au and $\nu=-63\pm21^{\circ}$.  \citet{fulle2016_67pdustdistribution} found dust-to-gas ratios (by mass), $f_{dg}$, between 5 and 10 for 67P/Churyumov-Gerasimenko, and so assuming a conservative value of $f_{dg}=5$ \citep[cf.][]{jewitt2016_324p}, this computed best-fit dust production rate for 238P corresponds to a water production rate of $Q_{\rm H_2O}\sim5\times10^{24}$ molecules~s$^{-1}$ (assuming water to be the dominant volatile material).

At the midpoint of the time period covered by this fitting analysis, water production rates are expected to range from ${\dot m}_w\sim3.0\times10^{-6}$ kg~s$^{-1}$~m$^{-2}$ in the isothermal (or ``fast-rotator'') approximation to ${\dot m}_w\sim5.6\times10^{-5}$ molecules~s$^{-1}$~m$^{-2}$ in the subsolar (or ``flat slab'') approximation for a sublimating graybody in thermal equilibrium (see Table~\ref{table:activity_fitting_results}).  Using
\begin{equation}
A_{\rm act} = {{\dot M}_d\over  f_{dg}{\dot m}_w}
\end{equation}
to determine the effective active area, $A_{\rm act}$, of a sublimating object, assuming $f_{dg}=5$, we find effective active area estimates ranging from $\sim3\times10^3$~m$^2$ (in the subsolar approximation) to $\sim5\times10^4$~m$^2$ (in the isothermal approximation).  Assuming the nucleus to be a spherical body with an effective radius of $r_N$ and using $r_N=400$~m as determined by \citet{hsieh2011_238p}, we then find effective active fraction estimates ranging from $f_{\rm act}\sim1\times10^{-3}$ to $f_{\rm act}\sim2\times10^{-2}$ (Table~\ref{table:activity_fitting_results}).

The best-fit pre-perihelion dust production rate that we find for 238P soon after it was confirmed to be active again in 2016 is a factor of a few larger than the average production rate of ${\dot M}_d$$\,=\,$0.2~kg~s$^{-1}$ determined for the object by \citet{hsieh2009_238p} from post-perihelion observations obtained in 2005-2007.  \citet{hsieh2009_238p} assumed a grain density of $\rho$$\,=\,$1000~kg~m$^{-3}$, though, whereas we assume $\rho$$\,=\,$2500~kg~m$^{-3}$ here, which could account for some of the difference in inferred dust production rates between the two epochs.  The available data also do not cover similar orbit arcs, with the data used for the analysis by \citet{hsieh2009_238p} having been obtained post-perihelion and at larger heliocentric distances, where temperatures and therefore sublimation rates would have been lower, than the pre-perihelion data analyzed here, making it difficult to definitively determine whether there have been changes in activity strength for 238P between 2005-2007 and 2016-2017.

On the other hand, data is available for the start of 238P's 2010-2011 active apparition \citep[Table~\ref{table:obs_238p_active_prev};][]{hsieh2011_238p}, providing an opportunity to perform a more direct comparison of activity strength for 238P during different active apparitions.  We follow the analysis applied above to photometric data from 238P's 2016 reactivation, fitting a linear function to estimated ejected dust masses for 238P computed from data obtained from 2010 September 3 to 2010 December 9, where the object's heliocentric distance change over this period corresponds to a $\sim$20\% (using the subsolar approximation) to 83\% (using the isothermal approximation) increase in the water sublimation rate on the object's surface.  Treating the lower limit excess mass measurements reported by \citet{hsieh2011_238p} as exact values (likely a reasonable approximation for this early period before dust in the comet's tail expands significantly beyond near-nucleus photometry apertures), we find a best-fit initial net dust production rate of ${\dot M}_d$$\,\sim\,$1.4$\pm$0.3~kg~s$^{-1}$ (corresponding to $Q_{\rm H_2O}\sim9\times10^{24}$ molecules~s$^{-1}$, assuming $f_{dg}=5$), about twice that found in our analysis of the object's 2016 activity.

We also find a best-fit start date for activity of $\sim$205$\pm$50~days prior to perihelion (corresponding to a best-fit start date of 2010 August 17 and an uncertainty range of 2010 June 28 to 2010 October 6), when the object was at a very similar heliocentric distance ($R=2.61\mp0.11$~au) and true anomaly ($\nu=-58\pm12^{\circ}$) as it was on the estimated start date of its 2016 activity.  We also find effective active area estimates ranging from $\sim6\times10^3$~m$^2$ (using the subsolar approximation) to $\sim1\times10^5$~m$^2$ (using the isothermal approximation), corresponding to effective active fraction estimates ranging from $\sim3\times10^{-3}$ (using the subsolar approximation) to $\sim7\times10^{-2}$ (using the isothermal approximation).  Additional details of our analysis of these data are summarized in Table~\ref{table:activity_fitting_results}.

Examining the true anomaly range over which we have overlapping data from 238P's 2010-2011 and 2016-2017 active periods ($-$35$^{\circ}$$\,<\,$$\nu$$\,<\,$$-$20$^{\circ}$), we find a excess dust mass of $M_d$$\,>\,$(1.5$\pm$0.5)$\times$10$^{7}$~kg on 2010 December 9 (when 238P was at $\nu$$\,=\,$$-$27.5$^{\circ}$; Table~\ref{table:obs_238p_active_prev}) and an average excess dust mass of $M_d$$\,=\,$(0.8$\pm$0.2)$\times$10$^{7}$~kg for the period between 2016 July 8 and 2016 August 6 (when 238P was at an average true anomaly of $\nu$$\,=\,$$-$27.3$^{\circ}$; Table~\ref{table:obs_238p_active}).  As such, we find that the excess dust mass present during the early portion of 238P's 2010-2011 active period may have been approximately twice that (or more) present during the same portion of 238P's 2016-2017 active period.

\subsection{288P Activity Characterization}\label{section:288p_active}

\begin{figure*}[htb!]
\centerline{\includegraphics[width=5.5in]{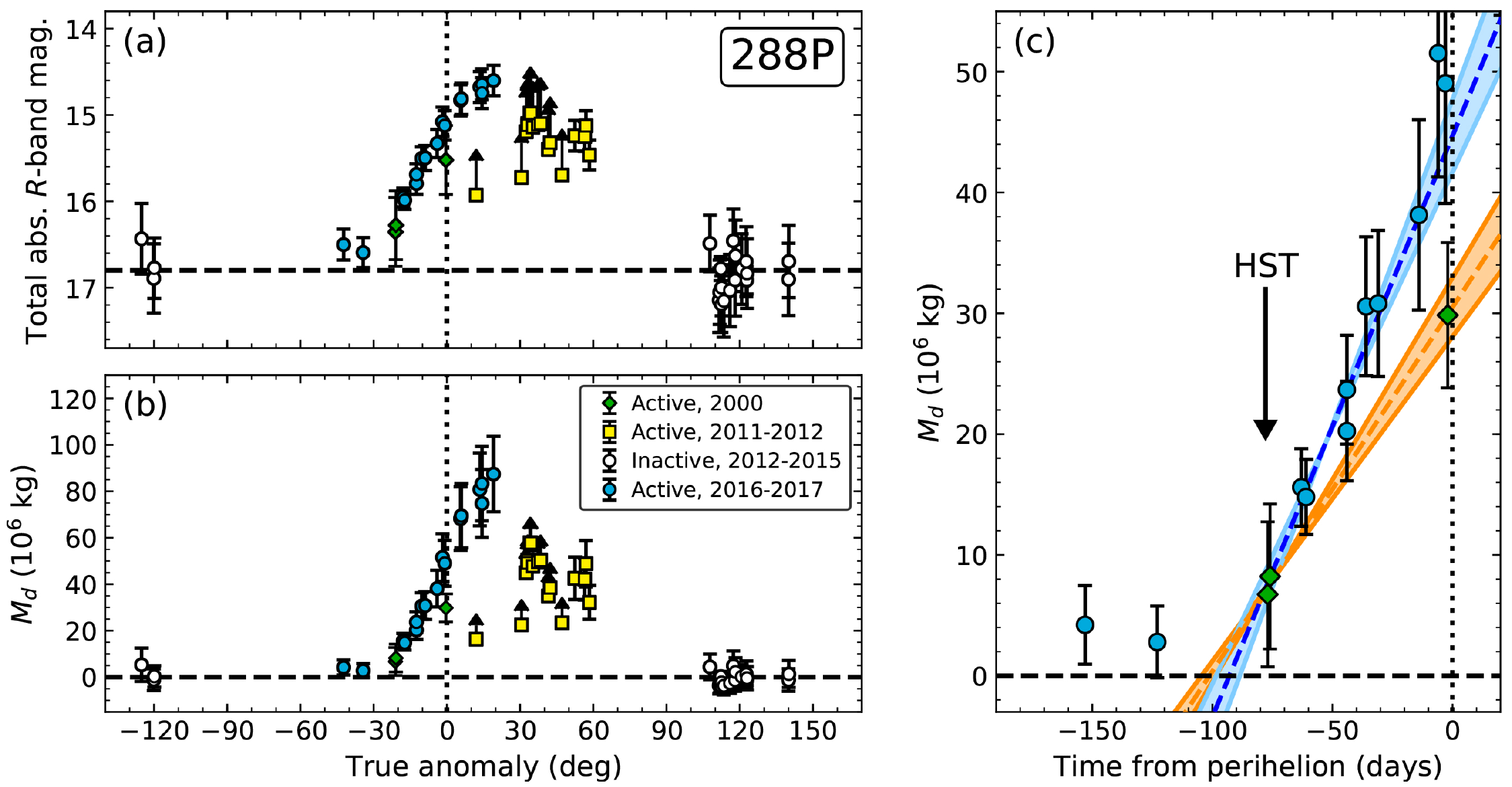}}
\caption{\small (a) Total absolute $R$-band magnitude of 288P during its 2000 active period (green diamonds), 2011-2012 active period (yellow squares), 2012-2015 inactive period (open circles), and 2016-2017 active period (blue circles) plotted as a function of true anomaly.  The expected magnitude of the inactive nucleus is marked with a horizontal dashed black line, while perihelion is marked with a dotted vertical line.  (b) Total estimated dust masses measured for 288P during the same periods of observations as in (a) plotted as a function of true anomaly.  The excess dust mass expected for the inactive nucleus (i.e., zero) is marked with a horizontal dashed black line, while perihelion is marked with a dotted vertical line.  (c)  Estimated total dust masses measured for 238P during just its 2016-2017 active period plotted as a function of time from perihelion (where negative values denote time before perihelion and positive values denote time after perihelion).  A diagonal solid blue line shows a linear fit to data obtained between 2016 June 8 and 2016 October 8 ($-42.1^{\circ}<\nu<-8.8^{\circ}$), reflecting the average net dust production rate over this period (over which dust production appears to be roughly linear) and allowing us to estimate the onset time of activity, while diagonal dotted blue lines show the range of uncertainty of the linear fit.  A diagonal solid green line shows a linear fit to data obtained between 2000 September 3 and 2000 November 17 ($-21.1^{\circ}<\nu<-0.4^{\circ}$), reflecting the average net dust production rate over this period (over which dust production appears to be roughly linear), while diagonal dotted green lines show the range of uncertainty of the linear fit.  A vertical arrow indicates the time of the HST observations of 288P by \citet{agarwal2016_288p_cbet} when the object was seen to be active.
}
\label{figure:images_288p_activityevolution}
\end{figure*}

To compare 288P's activity in 2016-2017 to its previously studied active apparition, we compute $M_d$ and $Af\rho$ for observations of 288P reported in \citet{hsieh2012_288p}.  Total magnitudes of 288P (i.e., including the entire coma and tail) were not measured for these data in the same way as we have measured recent data, and as such, we report estimated excess dust masses as lower limits in Table~\ref{table:obs_288p_active_prev}.  We plot total absolute $R$-band magnitudes and estimated ejected dust masses for 288P in Figure~\ref{figure:images_288p_activityevolution}.

Following the analysis performed above for 238P (Section~\ref{section:238p_active}), we fit a linear function to estimated ejected dust masses for 288P computed from data obtained between 2016 September 6 and 2016 October 25, where the object's heliocentric distance change over this period corresponds to a $\sim$13\% (using the subsolar approximation) to 50\% (using the isothermal approximation) increase in the water sublimation rate on the object's surface.
We find a best-fit initial net dust production rate for 288P early in its 2016-2017 active period of ${\dot M}_d$$\,=\,$5.6$\pm$0.7~kg~s$^{-1}$ (corresponding to $Q_{\rm H_2O}\sim4\times10^{25}$ molecules~s$^{-1}$, assuming $f_{dg}=5$), and a best-fit start date for activity of $\sim95\pm15$ days prior to perihelion (corresponding to a best-fit start date of 2016 August 5 and an uncertainty range of 2016 July 21 to 2016 August 20), when the object was at $R=2.48\mp0.02$~au and $\nu=-27\pm4^{\circ}$.
We also find effective active area estimates ranging from $\sim2\times10^4$~m$^2$ (using the subsolar approximation) to $\sim5\times10^5$~m$^2$ (using the isothermal approximation), corresponding to effective active fraction estimates ranging from $\sim1\times10^{-3}$ (using the subsolar approximation) to $\sim3\times10^{-2}$ (using the isothermal approximation).  These active fractions are calculated assuming that 288P's nucleus is a binary system with equally-sized spherical components, each with radii of $r_c\sim800$~m (Section~\ref{section:288p_inactive}). Additional details of our analysis of these data are summarized in Table~\ref{table:activity_fitting_results}.

While photometry of the object shows it to be slightly brighter than the expected magnitude of its inactive nucleus on 2016 June 8 and 2016 July 8, we choose to omit these data from the fitting analysis above as the photometric enhancements measured for the object on these dates are within the range of possible fluctuations in the object's brightness due to rotation, and as such, we cannot be certain that the object is in fact active on those dates.  As such, we omit these data points from this fitting analysis, focusing instead on data that we are certain was obtained when the object was active.

In their analysis of 288P's 2011-2012 active episode, \citet{licandro2013_288p} found an average dust production rate of 0.2~kg~s$^{-1}$ over a 100-day period starting shortly after perihelion (i.e., $0^{\circ}<\nu<30^{\circ}$; $2.43~{\rm au}<R<2.50~{\rm au}$), and a peak mass loss rate of 0.5~kg~s$^{-1}$ about 60 days after perihelion (when the object was at $\nu$$\,\sim\,$17$^{\circ}$ and $R$$\,\sim\,$2.46~au).  These conclusions were based on analysis of one night of data obtained on 2011 November 29 when the object was at $\nu$$\,=\,$37$^{\circ}$ and $R$$\,=\,$2.52~au.  \citet{licandro2013_288p} assumed a grain density of $\rho$$\,=\,$1000~kg~m$^{-3}$, whereas we assume $\rho$$\,=\,$2500~kg~m$^{-3}$ here, however, and also analyzed observations obtained at a larger heliocentric distance than the observations we discuss here.  Given the extremely large uncertainty on our computed mass loss rate and difference in initial assumptions, we regard our result as approximately consistent with that of \citet{licandro2013_288p}.

Although very little data is available for 288P from 2000 (Table~\ref{table:obs_288p_active}), we are interested in making at least a rough assessment of its activity strength at the time.
Following the analysis performed above, we fit a linear function to estimated ejected dust masses for 288P computed from data obtained between 2000 September 3 to 2000 November 17 (cf.\ Table~\ref{table:obs_288p_active_prev}), where the object's heliocentric distance change over this period corresponds to a $\sim$3\% (using the subsolar approximation) to 10\% (using the isothermal approximation) increase in the water sublimation rate on the object's surface.  We find a best-fit initial net dust production rate of ${\dot M}_d$$\,=\,$3.5$\pm$0.4~kg~s$^{-1}$ (corresponding to $Q_{\rm H_2O}\sim2\times10^{25}$ molecules~s$^{-1}$, assuming $f_{dg}=5$), and a best-fit start date for activity of $\sim100\pm15$~days prior to perihelion (corresponding to a best-fit start date of 2000 August 9 and an uncertainty range of 2000 July 25 to 2000 August 24), when the object was at $R=2.51\mp0.02$~au and $\nu=-28\pm4^{\circ}$.  We note that these uncertainties, which are generated from the same fitting routines used to analyze all of the other analogous data sets discussed in this work, imply a potentially unrealistically precise best-fit solution, given the small number of data points used to derive it, and in practice, should be regarded as being somewhat larger.  We also find effective active area estimates ranging from $\sim1\times10^4$~m$^2$ (using the subsolar approximation) to $\sim3\times10^5$~m$^2$ (using the isothermal approximation), corresponding to effective active fraction estimates ranging from $\sim9\times10^{-4}$ (using the subsolar approximation) to $\sim2\times10^{-2}$ (using the isothermal approximation).  Additional details of our analysis of these data are summarized in Table~\ref{table:activity_fitting_results}.  The estimated start time of 288P's activity in 2000 is quite close to its estimated start time in 2016, although the initial net dust production rate for the object appears to have actually increased from 2000 to 2016.

Examining the true anomaly range over which we have overlapping data from 288P's 2000 and 2016-2017 active periods, we find an average excess dust mass of $M_d$$\,=\,$(0.8$\pm$0.2)$\times$10$^7$~kg for 2000 September 3-4 when 288P had an average true anomaly of $\nu$$\,=\,$$-$21.0$^{\circ}$, and an average excess dust mass of $M_d$$\,=\,$(1.6$\pm$0.2)$\times$10$^7$~kg for 2016 September 6-8 when 288P had a similar average true anomaly of $\nu$$\,=\,$$-$17.5$^{\circ}$.  We also find an excess dust mass of $M_d$$\,=\,$(3.0$\pm$0.7)$\times$10$^7$~kg on 2000 November 17 when 288P was at $\nu$$\,=\,$$-$0.4$^{\circ}$, and an average excess dust mass of $M_d$$\,=\,$(5.1$\pm$0.7)$\times$10$^7$~kg for 2016 November 2-5 when 288P had a similar average true anomaly of $\nu$$\,=\,$$-$1.3$^{\circ}$.  As such, we find that excess dust masses measured for 288P in 2016 are larger than excess dust masses measured in 2000 for the object when it was at similar points in its orbit, corroborating our earlier conclusion that the average dust production rate for the object appears to have actually increased over time.

\subsection{Comparison to Other Active Asteroids}\label{section:comparison}

\begin{figure*}[htb!]
\centerline{\includegraphics[width=6.5in]{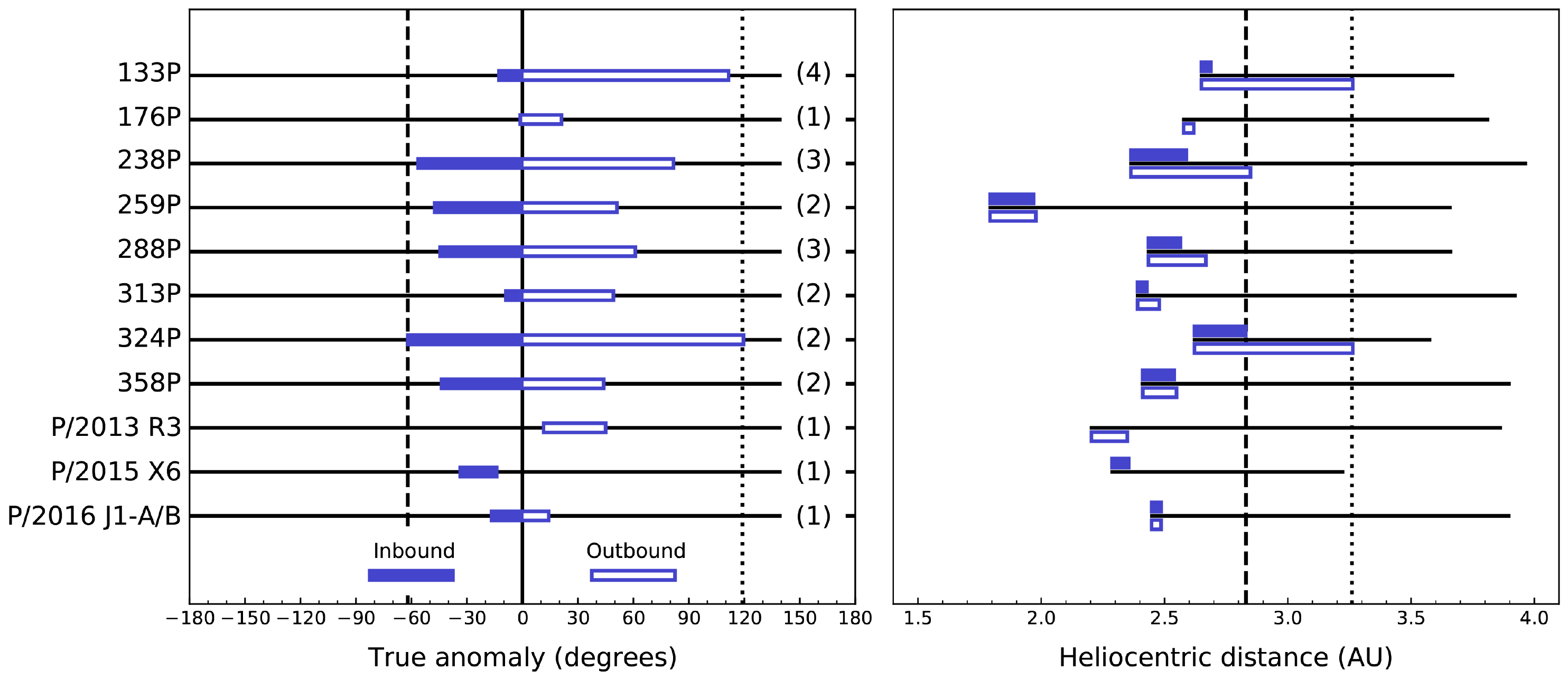}}
\caption{\small Active ranges, extending between the earliest and latest observations for which activity has been reported \citep[][and references within]{hsieh2015_ps1mbcs,moreno2016_p2015x6,hui2017_p2016j1,hsieh2017_259p}, in terms of true anomaly (left) and heliocentric distance (right) for likely MBCs. Solid blue line segments indicate the inbound (pre-perihelion) portion of each object's orbit while outlined blue line segments indicate the outbound (post-perihelion) portion of each object's orbit. In the left-hand panel, perihelion is marked with a dashed vertical line, while the earliest and latest orbit positions at which activity has been observed for 324P (the MBC with the earliest and latest observed activity in terms of true anomaly) are marked with dotted vertical lines. In the right-hand panel, horizontal black line segments indicate the heliocentric distance range covered by the orbit of each object, and the most distant positions at which activity has been observed for 324P (the MBC for which the most distant activity has been observed inbound to perihelion, as well as outbound away from perihelion) during the inbound and outbound portions of its orbit are marked with a vertical dashed line and a vertical dotted line, respectively. Numbers in parentheses to the far right of the left panel indicate the number of confirmed active apparitions that have been observed for each object.  After \citet{hsieh2015_324p}.
}
\label{figure:obs_all_activeasts}
\end{figure*}

We plot the confirmed active ranges (where visible dust emission or photometric enhancement has been reported) of all likely MBCs identified to date in Figure~\ref{figure:obs_all_activeasts}, updating the similar figure originally shown by \citet{hsieh2015_324p}.  While extending the active ranges confirmed for 238P and 288P based on observations presented here, we also add active ranges reported for recently discovered MBC candidates P/2015 X6 (PANSTARRS) and P/2016 J1-A/B (PANSTARRS) \citep[cf.][]{moreno2016_p2015x6,hui2017_p2016j1} to this figure, and extend the active range for 259P based on the recently reported confirmation of its reactivation \citep{hsieh2017_259p}.  As noted by \citet{hsieh2015_324p}, the active regions marked in this plot should be considered lower limits to the full ranges over which activity may be present for each object.  This is because good constraints are not always available for the onset or termination of activity for an object given that observational circumstances may prevent direct observations during the onset or termination of activity for objects already known to be active, or the fact that new active objects are by definition discovered while already exhibiting activity, and as such, must complete at least another full orbit before attempts can be made to directly observationally constrain the onset times of those objects' activity.

324P remains the MBC with the largest observed active range of all likely MBCs in terms of orbit position, with both the earliest and latest observations of activity in terms of true anomaly, and also the most distant confirmed activity in terms of heliocentric distance on both the inbound and outbound portions of its orbit of all of the MBCs.  However, the onset point of 238P's activity found by both \citet{hsieh2011_238p} and this work is similar to that of 324P in terms of true anomaly, while 133P has been observed to exhibit residual activity at a similarly large true anomaly and similarly distant heliocentric distance as 324P on the outbound portion of its orbit (cf.\ Figure~\ref{figure:obs_all_activeasts}).

\section{DISCUSSION}\label{section:discussion}

With reported observations of activity in 2016 and 2017 for both 238P and 288P and in 2000 for 288P \citep[this work;][]{agarwal2016_288p,hsieh2016_238p}, both objects have now been reported to be active near perihelion on three separate occasions.  This further solidifies the conclusion that their activity is likely to be due to sublimation of volatile material, and not due to disruptive events like rotational destabilization events or impacts, which would not be expected to repeat so regularly or so frequently, nor specifically occur near perihelion.

As discussed in Section~\ref{section:recurrent}, our observations of multiple active episodes are also useful for investigating the evolution of activity between those different active periods.  Interestingly, while activity strength (as parameterized by initial dust production rate) appears to have declined by about a factor of 2 for 238P between 2010-2011 and 2016-2017, activity strength for 288P appears to have increased between 2000 and 2016-2017.  Estimates of the evolution of water sublimation rates in the presence of a growing rubble mantle \citep[cf.][]{jewitt1996_dormantcomets} suggest that activity should decline relatively rapidly soon after activity is initially triggered but then should decline more slowly after a mantle of sufficient thickness has developed \citep[e.g.,][]{hsieh2015_ps1mbcs}.  This behavior can be partly attributed to the dependence of mantle growth rates on sublimation rates \citep[cf.][]{hsieh2015_ps1mbcs}: as sublimation rates decrease from mantling, mantle growth slows, slowing the change in sublimation rates from one orbit passage to the next.  A more detailed analysis by \citet{kossacki2012_259p} found that the evolution of activity for a MBC from one perihelion passage to the next also depend on additional physical properties and circumstances such as spin axis orientation, latitude of specific active sites, grain sizes, bulk density, and porosity.

Increases in activity strength over time could conceivably occur if other processes besides mantling also have significant modulating effects on dust production rates.  These processes may or may not be associated with the object's ongoing activity, and could include sinkhole collapses \citep[e.g.,][]{vincent2015_cometsinkholes} or rotation- or impact-induced landslide activity \citep[e.g.,][]{steckloff2016_hartley2avalanches,hofmann2017_impacttriggeredlandslides} that uncover fresh volatile material.  Perihelion distance reduction could also be a potential explanation for increased activity strength \citep{licandro2000_cometnuclei}.  In the case of 288P, further analysis (e.g., using detailed dust modeling) must first be done to confirm whether the apparent increase in activity strength between 2000 and 2016-2017 is real before speculating on possible or likely causes of such an increase.  Ultimately, determining whether physical processes that may actually increase activity strength over time are in operation on MBCs will likely require more occurrences of increasing activity strength to be identified and characterized.  To achieve this, detailed observational characterization of more repeated active episodes for MBCs will be required, further motivating continued monitoring of known active MBCs.



Meanwhile, we find similar start times for different active periods for both 238P (in 2010-2011 and 2016-2017) and 288P (in 2000 and 2016-2017) (Sections~\ref{section:238p_active} and \ref{section:288p_active}).  Assuming that each object's activity is driven by the sublimation of subsurface ice, the depth of that ice should be a significant controlling factor of when activity starts, given the finite time needed for solar insolation to propagate through surface layers to buried ice reservoirs \citep[cf.][]{hsieh2011_238p}.  As such, the consistent start times for activity for each object for different active periods appear to suggest that ice depths remained relatively consistent between the active episodes in question, implying that minimal mantle growth or ice recession occurred during these periods. 

In terms of follow-up opportunities, 238P was observable again from September 2017 to June 2018, during which it covered a true anomaly range of 80$^{\circ}$$\,<\,$$\nu$$\,<\,$130$^{\circ}$, while 288P was observable again from August 2017 to May 2018, during which it covered a true anomaly range of 70$^{\circ}$$\,<\,$$\nu$$\,<\,$125$^{\circ}$, offering opportunities to monitor both objects for activity past the largest true anomaly at which either 133P or 324P has been seen to exhibit activity.  The 2017-2018 observability window for 238P just barely overlapped the last observation of 238P's 2010-2011 active period that we report here, and so presented a potential opportunity for another direct comparison of activity strength for the object at the same point in its orbit during two different orbit passages in addition to the ones we present here.  The observing window for 288P did not overlap any previous active observations, however, and so no similar opportunity to directly compare activity strengths from two different orbit passages was available for this object.  We have acquired data for both objects during their respective follow-up periods and will report results based on analyses of those data in a future paper.

As indicated above in Section~\ref{section:recurrent} and in Figure~\ref{figure:obs_all_activeasts}, there are now seven MBCs which have been confirmed to exhibit activity on two or more separate occasions (always near perihelion, and often with intervening observational confirmation of inactivity away from perihelion). Of those, 238P and 288P have been seen to be active on three separate occasions and 133P has been seen to be active on four separate occasions.  Observations obtained during each of these active apparitions will be useful for systematic dust modeling studies, in which mass loss rates are independently determined for each active apparition in a consistent manner.  Results from these studies will allow us to more quantitatively compare changes in activity strength of MBCs from one active apparition to the next, giving us insights into the process of activity evolution over time for MBCs.  We plan to conduct such a study in the future, incorporating both data reported in the literature to date (including in this paper) and unpublished observations currently in hand or still being obtained of the recent reactivations of 259P, 324P, and 358P \citep[cf.][]{hsieh2015_324p,hsieh2017_259p,hsieh2018_358p}.  The number of currently active MBCs in the asteroid belt is related to the rate of triggering events (e.g., surface disruption events such as impacts or landslides) and the duration of activity following such triggering events \citep[][]{hsieh2009_htp}.  As such, given improved estimates of the size of the currently active MBC population provided by current and future survey data, better constraints on active lifetimes following triggering events enabled by an improved understanding of activity evolution could provide independent constraints on the rate of triggering events, providing a means for evaluating the plausibility of proposed triggering mechanisms.

\section{SUMMARY}\label{section:summary}
In this work, we present the following key findings:
\begin{enumerate}
\item{We confirm the reactivations of main-belt comets 238P/Read and 288P/(300163) 2006 VW$_{139}$, previously reported by \citet{hsieh2016_238p} and \citet{agarwal2016_288p_cbet}, and have obtained data following the evolution of each object's activity over several months in 2016 and 2017.  Additionally, we report the identification of archival SDSS data from 2000 of 288P in which the object is seen to be active.  With these observations, both 238P and 288P have now each been confirmed to be active near perihelion on three separate occasions.}
\item{A photometric analysis of observations obtained of 288P while the object appeared inactive from 2012 to 2015 yields best-fit IAU phase function parameters of $H_R$$\,=\,$16.80$\pm$0.12~mag and $G_R$$\,=\,$0.18$\pm$0.11, corresponding to an effective nucleus radius of $r_N=1.13\pm0.06$~km (assuming a $R$-band albedo of $p_R$$\,=\,$0.05) or, assuming 288P's nucleus to be a binary system with approximately equally sized components, effective component radii of $r_c=0.80\pm0.04$~km each.}
\item{In an analysis of our observations of 238P's reactivation in 2016, we find a best-fit initial average net dust production rate of ${\dot M_d}$$\,=\,$0.7$\pm$0.3~kg~s$^{-1}$ and a best-fit start date of activity of $\sim$225~days prior to perihelion, corresponding to 2016 March 11 when 238P was at $R=2.66$~au and $\nu=-63^{\circ}$.  In an analogous analysis of observations of 238P's activity in 2010-2011, we find a best-fit initial average net dust production rate of ${\dot M_d}$$\,\sim\,$1.4$\pm$0.3~kg~s$^{-1}$, i.e., about twice that estimated for the object's 2016-2017 active period, and a best-fit start date of activity of $\sim\,$205~days prior to perihelion, corresponding to 2010 August 17 when 238P was at $R=2.61$~au and $\nu=-58^{\circ}$, i.e., similar to the orbit position of the start of activity inferred for the object's 2016-2017 active period.  Comparing estimated dust masses from overlapping true anomaly ranges from 238P's 2010-2011 and 2016-2017 active periods, we find that the dust mass present in 2010-2011 may have been approximately twice that present over the same orbit arc in 2016-2017.}
\item{In an analysis of our observations of 288P's reactivation in 2016, we find a best-fit initial average net dust production rate of ${\dot M_d}=5.6\pm0.7$~kg~s$^{-1}$ and a best-fit start date of activity of $\sim\,$95~days prior to perihelion, corresponding to 2016 August 5 when 288P was at $R=2.48$~au and $\nu=-27^{\circ}$.  In an analogous analysis of our observations of 288P's reactivation in 2016, we find a best-fit initial average net dust production rate of ${\dot M_d}$$\,=\,$3.5$\pm$0.4~kg~s$^{-1}$, suggesting that the dust production rate actually increased between 2000 and 2016-2017, and a best-fit start date of activity of $\sim\,$100~days prior to perihelion, corresponding to 2000 August 9 when 288P was at $R=2.51$~au and $\nu=-28^{\circ}$, i.e., similar to the orbit position of the start of activity inferred for the object's 2016-2017 active period.  Comparing estimated dust masses from similar true anomaly positions during 288P's 2000 and 2016-2017 active periods, we find that excess dust masses estimated during 2016-2017 to be larger than those measured at similar orbit positions in 2000.  More detailed dust modeling and analysis will be required to determine whether the apparent increase in 288P's activity strength between 2000 and 2016-2017 is real, and if it is, what mechanisms could be responsible for such evolution of the object's activity strength.}
\item{We find similar start times for different active periods for both 238P (in 2010 and 2016) and 288P (in 2000 and 2016). The consistent start times for activity for each object for different active periods suggest that minimal mantle growth or ice recession occurred during the periods in question, leaving delays in the start of activity caused by the time needed for solar insolation to propagate through surface layers to buried ice reservoirs largely unchanged from one episode to the next.  We expect that future systematic dust modeling studies of the active apparitions of these objects and other MBCs will provide additional insights into the process of activity evolution for MBCs, with implications for constraining total activity lifetimes and the rate of MBC triggering events from discovery statistics.}
\end{enumerate}

\acknowledgments
HHH, MMK, NAM, and SSS acknowledge support from the NASA Solar System Observations program (Grant NNX16AD68G).  CAT was funded in part through the State
of Arizona Technology and Research Initiative Program.  HHH and MMK also thank the International Space Science Institute (ISSI) in Bern, Switzerland, for the hosting and provision of financial support for an international team to discuss the science of MBCs that facilitated discussion related to this work.

We are grateful to S.\ Arnouts, T. Burdullis, A.\ Draginda, N.\ Flagey, P.\ Forshay, A.\ Petric, L.\ Wells, and C.\ Wipper at CFHT, J.\ Ball, P.\ Candia, J.\ Chavez, D.\ Coulson, L.\ Fuhrman, M.\ Gomez, P.\ Hirst, M.\ Hoenig, J.\ Kemp, H.\ Lee, S.\ Leggett, A.\ Lopez, T.\ Matulonis, R.\ McDermid, J.\ Miller, C.\ Morley, J.\ O'Donoghue, S.\ Pakzad, M.\ Pohlen, J.\ Rhee, R.\ Salinas, D.\ Sanmartim, M.\ Schwamb, O.\ Smirnova, A.\ Smith, A.\ Stephens, S.\ Stewart, B.\ Walp, E.\ Wenderoth, and S.\ Yang at Gemini, and J.\ Dvorak, E.\ Moore, and M.\ Willman at the UH 2.2~m telescope for assistance in obtaining observations.
This work made use of the Discovery Channel Telescope operated by Lowell Observatory. Lowell is a private, non-profit institution dedicated to astrophysical research and public appreciation of astronomy and operates the DCT in partnership with Boston University, the University of Maryland, the University of Toledo, Northern Arizona University and Yale University.  The Large Monolithic Imager was built by Lowell Observatory using funds provided by the National Science Foundation (AST-1005313).

Funding for the SDSS and SDSS-II has been provided by the Alfred P.\ Sloan Foundation, the Participating Institutions, the National Science Foundation, the U.S.\ Department of Energy, the National Aeronautics and Space Administration, the Japanese Monbukagakusho, the Max Planck Society, and the Higher Education Funding Council for England. The SDSS Web Site is {\tt http://www.sdss.org/}.
The SDSS is managed by the Astrophysical Research Consortium for the Participating Institutions. The Participating Institutions are the American Museum of Natural History, Astrophysical Institute Potsdam, University of Basel, University of Cambridge, Case Western Reserve University, University of Chicago, Drexel University, Fermilab, the Institute for Advanced Study, the Japan Participation Group, Johns Hopkins University, the Joint Institute for Nuclear Astrophysics, the Kavli Institute for Particle Astrophysics and Cosmology, the Korean Scientist Group, the Chinese Academy of Sciences (LAMOST), Los Alamos National Laboratory, the Max-Planck-Institute for Astronomy (MPIA), the Max-Planck-Institute for Astrophysics (MPA), New Mexico State University, Ohio State University, University of Pittsburgh, University of Portsmouth, Princeton University, the United States Naval Observatory, and the University of Washington.
The Pan-STARRS1 Surveys (PS1) and the PS1 public science archive have been made possible through contributions by the Institute for Astronomy, the University of Hawaii, the Pan-STARRS Project Office, the Max-Planck Society and its participating institutes, the Max Planck Institute for Astronomy, Heidelberg and the Max Planck Institute for Extraterrestrial Physics, Garching, The Johns Hopkins University, Durham University, the University of Edinburgh, the Queen's University Belfast, the Harvard-Smithsonian Center for Astrophysics, the Las Cumbres Observatory Global Telescope Network Incorporated, the National Central University of Taiwan, the Space Telescope Science Institute, the National Aeronautics and Space Administration under Grant No. NNX08AR22G issued through the Planetary Science Division of the NASA Science Mission Directorate, the National Science Foundation Grant No.\ AST-1238877, the University of Maryland, Eotvos Lorand University (ELTE), the Los Alamos National Laboratory, and the Gordon and Betty Moore Foundation.

We wish to recognize and acknowledge the very significant cultural role and reverence that the summit of Maunakea has always had within the indigenous Hawaiian community.  We are fortunate to have the opportunity to conduct observations from this mountain.
        

%

\vspace{5mm}
\facilities{CFHT (MegaCam), Discovery Channel Telescope (LMI), Gemini North (GMOS-N), Gemini South (GMOS-S), Lulin One-meter Telescope, Magellan Baade (IMACS), University of Hawaii 2.2m, Sloan Digital Sky Survey (SDSS)}
\bibliographystyle{aasjournal}
\bibliography{hhsieh_refs}   




\end{document}